\documentstyle[PASJadd,epsf]{PASJ95}
\onecolumn
\begin{document}

\title{Nonlinear Stochastic Biasing of Peaks and Halos: \\
Scale-Dependence,  Time-Evolution, and  Redshift-Space Distortion \\
from Cosmological N-body Simulations}

\author{Atsushi {\sc Taruya}$^{1}$, Hiromitsu {\sc Magira}$^{1}$, 
Y.P. {\sc Jing}$^{2,3,4,5}$, and Yasushi {\sc Suto},$^{1,2}$ 
\\[12pt]
$^1$ {\it Department of Physics, School of Science, University of Tokyo, 
  Tokyo 113-0033}\\
{\it E-mail(AT): ataruya@utap.phys.s.u-tokyo.ac.jp}\\
$^2$ {\it Research Center for the Early Universe, School of Science, 
  University of Tokyo, Tokyo 113-0033} \\
$^3$ {\it Theory Division, National Astronomical Observatory, 
 Mitaka 181-8588}\\
$^4$ {\it Shanghai Astronomical Observatory, the Partner Group of MPI f\"ur Astrophysik,}\\{\it Nandan Road 80, Shanghai 20030, China}\\
$^5$ {\it National Astronomical Observatories, Chinese Academy of Sciences, Beijing 100012,China}\\
}

\abst{ We quantify the degree of nonlinearity and stochasticity of the
  clustering of biased objects, using cosmological N-body
  simulations. Adopting the peaks and the halos as representative
  biasing models, we focus on the two-point correlation of the biased
  objects, dark matter and their cross-correlation. Especially, we
  take account of the effect of redshift-space distortion and attempt
  to clarify the scale-dependence and the time-dependence by analyzing
  the biasing factor and the cross-correlation factor. On small
  scales, stochasticity and nonlinearity become appreciable and
  strongly object-dependent, especially in redshift space due to the
  pair-wise velocity dispersion of the biased objects.  Nevertheless,
  an approximation of deterministic linear biasing $\delta_{\rm
  \scriptscriptstyle obj}\simeq b_{\rm \scriptscriptstyle obj}
  \delta_{\rm \scriptscriptstyle mass}$ works reasonably well even in
  the quasi-linear regime $r \lower.5ex\hbox{$\; \buildrel > \over
  \sim \;$} 10 h^{-1}$ Mpc, and linear redshift-space distortion
  explains the clustering amplitudes in redshift space in this
  regime. }

\kword{ Cosmology: dark matter - Galaxies:clustering - halos -
  Large-scale structure of universe - Theory}

\maketitle

\thispagestyle{headings}

\section{Introduction}

Luminous objects, such as galaxies and quasars, are not direct tracers
of the mass in the universe. In fact, the difference of the spatial
distribution between luminous objects and dark matter, or the {\it
bias}, has been indicated from a variety of observations.  In order to
confront theoretical model predictions for the {\it mass} distribution
against observational data, one needs a relation of density fields of
mass and luminous objects. Consider the density contrasts of visible
objects and mass, $\delta_{\rm \scriptscriptstyle obj}({\bf x},z|R)$
and $\delta_{\rm \scriptscriptstyle mass}({\bf x},z|R)$, at a position
${\bf x}$ and a redshift $z$ smoothed over a scale $R$. In general,
the former should depend on various other auxiliary variables
$\vec{\cal A}$ defined at different locations ${\bf x'}$ and redshifts
$z'$ smoothed over different scales $R'$ in addition to the mass
density contrast at the same position, $\delta_{\rm \scriptscriptstyle
mass}({\bf x},z|R)$.  While this relation can be schematically
expressed as
\begin{equation}
  \label{eq:generalbias}
  \delta_{\rm \scriptscriptstyle obj}({\bf x},z|R)=
{\cal F}[{\bf x},z, R, 
  \delta_{\rm \scriptscriptstyle mass}({\bf x},z|R),
\vec{\cal A}({\bf x'},z'|R'), \ldots ] ,
\end{equation}
it is impossible even to specify the list of the astrophysical
variables $\vec{\cal A}$, and thus hopeless to predict the functional
form in a rigorous manner. Therefore if one simply focuses on the
relation between $\delta_{\rm \scriptscriptstyle obj}({\bf x},z|R)$
and $\delta_{\rm \scriptscriptstyle mass}({\bf x},z|R)$, the relation
becomes inevitably {\it stochastic} and {\it nonlinear} due to the 
dependence on unspecified auxiliary variables $\vec{\cal A}$. 

For illustrative purposes, define the {\it biasing} factor as the
ratio of the density contrasts of luminous objects and mass:
\begin{equation}
  \label{eq:bgeneral}
  B_{\rm \scriptscriptstyle obj}({\bf x},z|R)
\equiv \frac{\delta_{\rm \scriptscriptstyle obj}({\bf x},z|R)}
{\delta_{\rm \scriptscriptstyle mass}({\bf x},z|R)}
= \frac
{{\cal F}[{\bf x},z, R, 
  \delta_{\rm \scriptscriptstyle mass}({\bf x},z|R),
\vec{\cal A}({\bf x'},z'|R'), \ldots ] }
{\delta_{\rm \scriptscriptstyle mass}({\bf x},z|R)} .
\end{equation}
Only in very idealized situations, the above {\it nonlocal stochastic
  nonlinear} factor in terms of $\delta_{\rm \scriptscriptstyle mass}$
may be approximated by
\begin{enumerate}
\item a {\it local stochastic nonlinear} bias:
\begin{equation}
  \label{eq:localbias}
  B_{\rm \scriptscriptstyle obj}({\bf x},z|R)=
b_{\rm \scriptscriptstyle obj}^{\rm \scriptscriptstyle (sn)}[{\bf x},z, R, 
  \delta_{\rm \scriptscriptstyle mass}({\bf x},z|R),
\vec{\cal A}({\bf x},z|R), \ldots ] ,
\end{equation}
\item a {\it local deterministic nonlinear} bias:
\begin{equation}
  \label{eq:ldbias}
  B_{\rm \scriptscriptstyle obj}({\bf x},z|R)=
b_{\rm \scriptscriptstyle obj}^{\rm \scriptscriptstyle (dn)}[z, R, 
  \delta_{\rm \scriptscriptstyle mass}({\bf x},z|R)] ,
\end{equation}
and
\item a {\it local deterministic linear} bias:
\begin{equation}
  \label{eq:linearbias}
  B_{\rm \scriptscriptstyle obj}({\bf x},z|R) 
= b_{\rm \scriptscriptstyle obj}(z, R) 
\end{equation}
\end{enumerate}

From the above point of view, the local deterministic linear bias is
obviously unrealistic, but is still a widely used conventional model
for biasing. In fact, the time- and scale-dependence of the linear
bias factor $b{\rm \scriptscriptstyle obj}(z, R)$ was neglected in
many previous studies of biased galaxy formation until very
recently. Currently, however, various models beyond the deterministic linear
biasing have been seriously considered with particular emphasis on the
nonlinear and stochastic aspects of the biasing (Pen 1998; Tegmark,
Peebles 1998; Dekel, Lahav 1999; Taruya, Koyama, Soda 1999; Tegmark,
Bromley 1999; Blanton et al. 1999; Somerville et al. 1999; Blanton et
al. 2000; Taruya 2000; Taruya, Suto 2000; Yoshikawa et al. 2000).

In the present paper, we examine two fairly simplified but
well-defined popular models for biasing, i.e., peaks and halos, paying
particular attention to their stochastic and nonlinear nature on the
basis of cosmological N-body simulations. Especially we take account
of the redshift-space distortion effect on biasing which has not been
studied in previous work. Our main purpose is to find the validity and
limitation of the local deterministic linear approximation
$\delta_{\rm \scriptscriptstyle obj}= b_{\rm \scriptscriptstyle
obj}\delta_{\rm \scriptscriptstyle mass}$ for the general stochastic
and nonlinear biasing.  By analyzing the two-point statistics of the
biased objects and the mass distribution, we quantitatively measure
the stochasticity and nonlinearity and investigate the
scale-dependence and the time-evolution of clustering amplitude both
in real and redshift spaces.

The nonlinear stochastic biasing for two-point statistics is not a
simple theoretical issue, but important in the proper comparison with
observation.  Although the biasing properties for one-point statistics
have been extensively discussed in literature using numerical
simulations and/or analytic models, a similar investigation of biasing
in two-point statistics has not yet performed. This is because
nonlinear stochastic biasing was originally formulated using the
one-point statistics.  In this paper, we quantify the biasing features
of clustering in terms of the two-point correlation function. In
particular, we focus on the biasing factor $b_{\rm\scriptscriptstyle
obj}$ and the cross correlation coefficient $r_{\rm\scriptscriptstyle
obj}$ defined by equations.(\ref{eq:def_b}) and
(\ref{eq:def_r}). Those quantities are related and compared to the
biasing parameters defined in terms of the one-point statistics.

The paper is organized as follows. In section 2, we briefly describe
the simulated catalogues and evaluate the two-point correlation
functions for the peaks and dark halos. We then examine the
scale-dependence and the time-evolution of biasing measured from those
statistics in section 3. In particular we pay attention to the
difference of the biasing properties of peaks and halos.  The results
are compared with simple theoretical models and predictions from
linear theory of redshift-space distortion.  In addition, we calculate
the variances of peaks, halos and dark matter particles taking into
account the point process.  Section 4 is devoted to the conclusions
and discussion.

\section{Peak and halo catalogues from cosmological N-body simulations}
%
\subsection{Simulations}

In this paper, we use a series of cosmological N-body simulations in
cold dark matter (CDM) cosmogonies (Jing, Suto 1998), whose
parameters are listed in Table 1.  All the models employ $N=256^3$
dark matter particles in the periodic comoving cube of the boxsize
$L_{\rm box} = 300h^{-1}$Mpc, and are evolved on the basis of the
Particle-Particle -- Particle-Mesh (P$^3$M) method.  The initial
conditions of the particle distribution match the CDM transfer
function of Bardeen et al. (1986) characterized by the shape
parameter, $\Gamma$. The RMS mass fluctuation amplitude at $8
h^{-1}$Mpc, $\sigma_8$, is normalized according to the cluster
abundance (Kitayama, Suto 1997).  We analyze three realizations for
each cosmological model at $z=0$ and $z=2.2$, and one realization for
LCDM model at $z=0.6$, $1.0$, and $z=3.4$ to examine the time
evolution.

As specific models of biasing, we consider two popular cases;
primordial density peaks and dark matter halos. The density peaks are
selected according to the algorithm of Mo, Jing \& White (1997). The
initial density field is smoothed with a Gaussian window. The
filtering radius of $R_{\rm f}= 0.54\,h^{-1}\,{\rm Mpc}$ is adopted
for the smoothing so that the typical size of the peaks becomes
relevant for galactic-sized objects (Bardeen et al. 1986). Given a
density field and a threshold peak height, this algorithm predicts the
number of the peaks that each simulation dark matter particle will
carry. The number of peaks per dark matter particle in our simulations
is always less than 1.  So we have picked up the peaks by randomly
selecting simulation particles with the selection probability set
equal to its number of the peaks for each particle. This way
guarantees that a correct total number of peaks is reproduced for the
Gaussian random field.  Those peak particles subsequently move
according to the gravitational field, and thus their total number is
conserved.  We choose $\nu_{\rm th}=1.0$, 2.0, and 3.0 as the
threshold of the peak height $\nu$ (Table 2).  As for the dark matter
halos, these are identified using the standard friend-of-friend
algorithm with a linking length of 0.2 in units of the mean particle
separation. We select halos of mass larger than the threshold $M_{\rm
th}=2.0\times 10^{12}$, $5.0\times 10^{12}$ and $1.0\times
10^{13}\,h^{-1}\,M_{\odot}$ (Table 3). The halos having the dark
matter particle $N_{\rm mass}<10$ have been excluded.

Figure 1 depicts the distribution of dark matter particles ({\it
upper-panel}), peaks ({\it middle-panel}) and halos ({\it lower-panel})
in LCDM model at $z=0$ and $z=2.2$ within a circular slice ({\it
comoving} radius of $150h^{-1}$Mpc and thickness of $15h^{-1}$Mpc).  We
locate a fiducial observer in the center of the circle. Then the {\it
comoving} position vector {\bf r} for a particle with a {\it comoving}
peculiar velocity {\bf v} at a redshift $z$ is observed at the position
{\it s} in redshift space:
\begin{equation}
  \label{eq:s-r}
  {\bf s}={\bf r}+\frac{1}{H(z)}\, \frac{{\bf r}\cdot{\bf v}}{|\bf r|}
\, \frac{{\bf r}}{|\bf r|} ,
\end{equation}
where $H(z)$ is the Hubble parameter at $z$.  The right panels in
Figure 1 plot the observed distribution in redshift space, where the
redshift-space distortion is quite visible; the coherent velocity
field enhances the structure perpendicular to the line-of-sight of the
observer ({\it squashing}) while the virialized clump becomes
elongated along the line-of-sight ({\it finger-of-God}).  Although the
distribution of peaks in Figure 1a and 1b seems rather different, this
is simply due to the strong concentration along the filamentary
regions in Figure 1a, and the total number of peaks is the same.
Figure 1 qualitatively illustrates that the redshift-space distortion
is sensitively dependent on the objects, and this is what we will
attempt to quantify statistically in the rest of paper.

\subsection{Two-point statistics}

We use the two-point correlation functions to quantify stochasticity
and nonlinearity in biasing of peaks and halos, and explore the
signature of the redshift-space distortion. Since we are interested in
the relation of the biased objects and the dark matter, we introduce
three correlation functions; the auto-correlation functions of dark
matter and the objects, $\xi_{\rm \scriptscriptstyle mm}$ and
$\xi_{\rm \scriptscriptstyle oo}$, and their cross-correlation function
$\xi_{\rm \scriptscriptstyle om}$. In the present case, the subscript
o refers to either h (halos) or $\nu$ (peaks). We also use the
superscripts R and S to distinguish quantities defined in real and
redshift spaces, respectively.  We estimate those correlation
functions using the standard pair-count method.  The correlation
function $\xi^{(S)}$ is evaluated under the distant-observer
approximation.

Those correlation functions are plotted in Figures 2 and 3 for peaks
and halos, respectively. Their qualitative behavior in redshift space
is already well-known and easy to understand (e.g, Kaiser 1987;
Hamilton 1998).  The correlation functions of biased objects generally
are larger than those of mass amplitudes. In nonlinear regimes ($\xi >
1$) the finger-of-god effect suppresses the amplitude of $\xi^{\rm
(S)}$ relative to $\xi^{\rm (R)}$, while $\xi^{\rm (S)}$ is larger
than $\xi^{\rm (R)}$ in linear regimes ($\xi < 1$) due to the coherent
velocity field.  In the next section, we present more quantitative
analysis in terms of the stochastic nonlinear biasing scheme and
compare with the existing theoretical predictions valid in linear
regimes.

\section{Scale-dependence and time-evolution of biasing parameters}

In order to quantify the behavior of biasing, we introduce the biasing
factor and the cross-correlation factor:
\begin{eqnarray}
\label{eq:def_b}
  b_{\rm \scriptscriptstyle obj}(r,z) 
&\equiv& \sqrt{\frac{\xi_{\rm oo}}{\xi_{\rm mm}}} , \\
\label{eq:def_r}
  r_{\rm \scriptscriptstyle obj}(r,z)
&\equiv&  \frac{\xi_{\rm mo}}{\sqrt{\xi_{\rm oo}\xi_{\rm mm}}} , 
\end{eqnarray}
following their counterparts defined in terms of the one-point
statistics (e.g., Pen 1998; Tegmark, Peebles 1998; Dekel, Lahav 1999;
Taruya, Koyama, Soda 1999; Taruya 2000).  
In equations (\ref{eq:def_b}) and (\ref{eq:def_r}), 
$b_{\rm \scriptscriptstyle obj}$ measures the relative strength of the
clustering of objects and $r_{\rm \scriptscriptstyle obj}$ 
characterizes the degree of nonlinearity and stochasticity of the underlying 
biasing mechanism (Dekel, Lahav 1999).  The
deterministic linear bias $\delta_{\rm \scriptscriptstyle obj}=b_{\rm
  \scriptscriptstyle obj}\delta_{\rm \scriptscriptstyle mass}$, for
instance, implies that $r_{\rm \scriptscriptstyle obj}=1$. 

Since the two-point correlation functions are not positive definite,
however, the physical interpretations of the above definitions are
somewhat ambiguous in that case. We will discuss this point in detail
in Sec.3.4 where we relate them to those defined by the one-point
statistics, i.e, $b_{\rm \scriptscriptstyle obj}^{(1)}$ and $r_{\rm
\scriptscriptstyle obj}^{(1)}$.  In any case, this does not happen in
our analysis below on scales of cosmological interest, and we adopt
the above ``conventional'' definitions.  In what follows, we focus on
the scale-dependence, time-evolution of the above parameters, as well
as on their behavior in redshift space. Before presenting the
simulation results, we briefly summarize the existing theoretical
models. They turn out to be useful in comparing and understanding our
results despite that they are valid mainly in linear regimes.

\subsection{Theoretical models for biasing}

\subsubsection{Peak model}

Density peaks are the first physical model of cosmological biasing
proposed by Kaiser (1984), and then extensively studied by Bardeen et
al. (1986). According to the latter, we consider the density peaks in
the {\it primordial} random Gaussian field smoothed with a Gaussian
window function. Then the height of each peak is defined as the ratio
of the density contrast at the position smoothed with $R_{\rm f}$ and
its RMS value, $\nu=\delta/\sigma_0$.  Although $\delta$ and
$\sigma_{0}$ in this context should refer to the values at some
initial epoch, we take another (but equivalent) view that their
values are linearly extrapolated to the present time. To be more
specific, $\sigma_0$ is computed by setting $l=0$ in the following
general expression for the $l$-th order moment:
\begin{eqnarray}
    \sigma_{l}^{2}(R_{\rm f})
       \equiv \int \frac{d^3k}{(2\pi)^{3}}
       \, k^{2l} \, P(k;z=0)\, \exp\left(-k^2R_{\rm f}^2\right) ,
\end{eqnarray}
where $P(k;z=0)$ is the linearly extrapolated power spectrum.

Bardeen et al. (1986) showed that the number density of peaks, $n_{\rm
\scriptscriptstyle peak}$, is given by
\begin{equation}
  \label{eq:n_peak} 
n_{\rm \scriptscriptstyle peak}
  (\nu) = \frac{1}{(2\pi)^{2}R_{*}^{3}} e^{-\nu^{2}/2}
  G(\gamma,\gamma\nu) ,
\end{equation}
where 
\begin{eqnarray}
R_{*} &=& \sqrt{3} \frac{\sigma_1(R_{\rm f})}{\sigma_2(R_{\rm f})} ,
\\
\gamma &=& \frac{\sigma_1^2(R_{\rm f})} 
{\sigma_0(R_{\rm f})\sigma_2(R_{\rm f})} , 
\\
G(\gamma,y) &=& \int_0^\infty dx \, F(x) \,
  \frac{\exp[-(x-y)^2/2(1-\gamma^2)]}{[2\pi(1-\gamma^2)]^{1/2}} , 
\\
F(x) &=&\frac{x^3-3x}{2}\left\{{\rm erf}\left[\sqrt{\frac{5}{2}}\,x\right]
      +{\rm erf}\left[\sqrt{\frac{5}{2}}\,\frac{x}{2}\right]\right\} 
    + \sqrt{\frac{2}{5\pi}}\left[\left(\frac{31}{4}x^{2}+
        \frac85\right)e^{-5x^2/8}
      +\left(\frac{x^2}{2}-\frac{8}{5}\right)e^{-5x^2/2}\right] .
\end{eqnarray}

Mo, Jing \& White (1997) derived a formula which relates the number
density fluctuations of peaks to the mass density fluctuations
$\delta_{\rm mass}$ at $z$.  Denoting $D(z)$ as the linear growth
factor normalized to unity at $z=0$, the extrapolated amplitude of
mass fluctuations at the present time is $\delta_{\rm
mass}/D(z)$. Then, the fluctuation $\delta_{\rm \scriptscriptstyle
peak}$ with $\nu \gg \delta_{\rm \scriptscriptstyle
mass}/[D(z)\sigma_0]$ is given by
\begin{eqnarray}
\label{eq:delta_peak}
  \delta_{\rm \scriptscriptstyle peak}(\nu,z) = 
(1+\delta_{\rm \scriptscriptstyle mass})
  \frac{n_{\rm \scriptscriptstyle peak}(\nu)|_
  {\nu \to \nu-\delta_{\rm \scriptscriptstyle mass}/[D(z)\sigma_0]} }
  {n_{\rm \scriptscriptstyle peak}(\nu)} -1
  \simeq \left\{1- \frac{1}{D(z)\sigma_0}\frac{d}{d\nu}\log
          \left[n_{\rm \scriptscriptstyle peak}(\nu)\right]
  \right\}\delta_{\rm \scriptscriptstyle mass}, 
\end{eqnarray}
in the limit of linear theory. Thus the peak model reduces to 
the scale-independent deterministic linear biasing:
\begin{eqnarray}
  \label{eq:b_peak}
  b_{\rm \scriptscriptstyle peak}(\nu,z)
&\equiv& \frac{\delta_{\rm \scriptscriptstyle peak}}
{\delta_{\rm \scriptscriptstyle mass}}
  = 1+\frac{1}{D(z)}\,\,\frac{\nu^2+g_1}{\sigma_0(R_{\rm f})\nu} ,
\\
  g_1 &\equiv& -\frac{\gamma\nu}{G(\gamma,\gamma\nu)}
  \left. \frac{\partial G(\gamma,y)}{\partial y}
\right|_{y=\gamma\nu}  . 
\end{eqnarray}
Since our simulation data are for peaks with the peak height above
$\nu$, we use the following effective biasing factor in the comparison
below:
\begin{equation}
  \label{eq:biaspeakeff}
  b_{\rm \scriptscriptstyle peak,eff}(>\nu,z)=
    \frac{\displaystyle  \int_\nu^\infty 
    b_{\rm \scriptscriptstyle peak}(\nu',z)
    n_{\rm \scriptscriptstyle peak}(\nu')d\nu'}
{\displaystyle \int_\nu^\infty 
    n_{\rm \scriptscriptstyle peak}(\nu')d\nu'} .
\end{equation}

\subsubsection{Halo biasing (Merging model)}

Another popular model of cosmological biasing is that one dark matter
halo accommodate a single luminous object. This is established as a
fairly realistic model for clusters of galaxies (e.g, Kitayama, Suto
1996), and may provide a reasonable approximation even for bright
galaxies (Somerville et al. 2000).  The number density of halos with
mass $M$ at $z$ is well approximated by the Press--Schechter mass
function (Press, Schechter 1974):
\begin{eqnarray}
  \label{eq:n_halo}
  n_{\rm \scriptscriptstyle halo}(M,z)dM &=& 
  \frac{1}{\sqrt{2\pi}}\frac{\bar{\rho}}{M}
  \frac{\delta_{c}}{\Delta^{3}(M,z)}
  \exp\left[-\frac{\delta_{c}^{2}}{\Delta^{2}(M,z)}\right]
  \left|\frac{d\Delta^{2}(M,z)}{dM}\right|dM , \\
\label{eq: mass_var}
   \Delta^{2}(M,z)
       &=& D^2(z) \int \,\frac{d^3k}{(2\pi)^{3}}\,
       P(k;z=0) \,\, \frac{9}{(kR_{\rm\scriptscriptstyle TH})^{6}}
 \left[\sin (kR_{\rm\scriptscriptstyle TH})
-kR_{\rm\scriptscriptstyle TH}
\cos (kR_{\rm\scriptscriptstyle TH})\right]^2, 
\end{eqnarray}
where $\bar{\rho}$ is the mean mass density, $\delta_c$ is the
critical density threshold of the spherical collapse, and the
spherical top-hat radius $R_{\rm\scriptscriptstyle TH}$ is
$(3M/4\pi\bar{\rho})^{1/3}$.

An analytic model for dark halo biasing is constructed in a similar way
as the peak model (Mo, White 1996; Mo et al. 1997), yielding
\begin{eqnarray}
  \delta_{\rm \scriptscriptstyle halo}(M,z) =
   (1+\delta_{\rm \scriptscriptstyle mass})
  \frac{n_{\rm \scriptscriptstyle halo}(M,z)
  |_{\delta_{c}\to\delta_{c}-\delta_{\rm \scriptscriptstyle mass}}}
  {n_{\rm \scriptscriptstyle halo}(M,z)} -1
  \simeq \left\{1-\frac{d}{d\delta_{c}}\log
          \left[n_{\rm \scriptscriptstyle halo}(M,z)\right]
  \right\}\delta_{\rm \scriptscriptstyle mass} .
\end{eqnarray}
Again in the above limit of linear regime, the halo biasing becomes
linear, deterministic and scale-independent.  In what follows, we
adopt the fitting formula of Jing (1998) for $b_{\rm
\scriptscriptstyle halo}(M,z)$ which incorporates a correction factor
for low mass halos:
\begin{eqnarray}
\label{eq:b_halo}
    b_{\rm \scriptscriptstyle halo}(M,z) 
&=& \left(\frac{0.5}{\nu_{\rm h}^4}+1\right)^{(0.06-0.02n_{\rm eff})}
    \left(1+\frac{\nu_{\rm h}^2-1}{\delta_{c}}\right) , \\
  \nu_{\rm h} &=& \frac{\delta_{c}}{\Delta(M,z)} , \\
  n_{\rm eff} &=&  \left. \frac{d\ln P(k)}{d\ln k}\right|_{k=2\pi/R} ,
\end{eqnarray}
and introduce the effective halo biasing factor to compare with the
simulation results:
\begin{equation}
\label{eq:biashaloeff}
  b_{\rm \scriptscriptstyle halo,eff}(>M,z) = 
\frac{\displaystyle 
\int_{M}^\infty b_{\rm \scriptscriptstyle halo}(M',z) \, 
n_{\rm \scriptscriptstyle halo}(M',z)dM'}
{\displaystyle \int_{M}^\infty n_{\rm\scriptscriptstyle halo}(M',z)dM'} .
\end{equation}

\subsubsection{Time-evolution of biasing parameters 
in a number-conserving model}

As noted above, both peak and halo models result in linear,
deterministic and scale-independent biasing on large scales. In this
limit, time-evolution of the biasing parameters can be generally
computed using the continuity equation if the number of objects is
constant (number conserving model). This is indeed the case for the
peak model, while not for the halo model.

The explicit expression for the time-dependent biasing factor is
obtained first by Fry (1996) in a context of the deterministic biasing
model. This is later extended to the stochastic biasing case in the
linear regime by Tegmark \& Peebles (1998), and then in the weakly
nonlinear regime by Taruya, Koyama \& Soda (1999) and Taruya
(2000). We use the expression of Tegmark \& Peebles (1998) in
considering the evolution of peak bias, which relates ($b_{\rm
\scriptscriptstyle obj}$, $r_{\rm \scriptscriptstyle obj}$) at a given
redshift $z_1$ to those at an earlier epoch, $z_{2}(>z_1)$:
\begin{eqnarray}
  \label{eq:tp98bias}
  b_{\rm \scriptscriptstyle obj}(z_1) &=& 
  \frac{\sqrt{D^2(z_2,z_1)-2D(z_2)
  b_{\rm \scriptscriptstyle obj}(z_2) 
r_{\rm \scriptscriptstyle obj}(z_2)D(z_2,z_1)
   +D^2(z_2)b_{\rm \scriptscriptstyle obj}(z_2)^2}}{D(z_1)} , \\
  \label{eq:tp98coe}
  r_{\rm \scriptscriptstyle obj}(z_1) &=& 
  \frac{D(z_2)b_{\rm \scriptscriptstyle obj}(z_2)
    r_{\rm \scriptscriptstyle obj}(z_2) - D(z_2,z_1)}
    {D(z_1)b_{\rm \scriptscriptstyle obj}(z_1)}, 
\end{eqnarray}
where we define
\begin{equation}
  \label{eq:dz2z1}
  D(z_2,z_1) \equiv D(z_2)-D(z_1).
\end{equation}
We note that in the case of a deterministic linear bias, $r_{\rm
  \scriptscriptstyle obj}$ becomes unity and the above expression
reduces to
\begin{eqnarray}
 b_{\rm \scriptscriptstyle obj}(z_1)=1+\frac{D(z_2)}{D(z_1)} 
\left[b_{\rm \scriptscriptstyle obj}(z_2)-1\right] ,
\end{eqnarray}
which is identical to the behavior of $b_{\rm \scriptscriptstyle
peak}$ (eq.[\ref{eq:b_peak}]) in a linear regime.

\subsection{Scale-dependence}

With the above theoretical predictions as a reference, let us examine
the scale-dependence of the biasing parameters.  Figures 4 and 5
display the results for peak and halo models, respectively.  Apart
from the object-dependent amplitude of the parameters, the general
trend is quite similar; on large scales, $r^{(R)}_{\rm
\scriptscriptstyle obj}$ is close to unity and the deterministic
linear bias is a reasonable approximation in this linear regime. In
fact, both $b^{(R)}_{\rm \scriptscriptstyle peak}$ and $b^{(R)}_{\rm
\scriptscriptstyle halo}$ approach the predictions
(\ref{eq:biaspeakeff}) and (\ref{eq:biashaloeff}), respectively, as
the separation $r$ increases.  On smaller scales, the biasing becomes
nonlinear and stochastic (i.e., $r^{(R)}_{\rm \scriptscriptstyle obj}
<1$) and the clustering amplitude relative to mass becomes generally
stronger ($b^{(R)}_{\rm \scriptscriptstyle obj}$ increases). Although
$b^{(R)}_{\rm \scriptscriptstyle halo}$ levels off and then {\it
decreases} below a few Mpc, this should be mainly ascribed to the
exclusion effect due to the finite halo size (Mo, Jing, White 1997).

Turn next to the redshift-space distortion effect on the
scale-dependence.  On large scales, the simulation results are
understood fairly reasonably with a straightforward extension of a
linear distortion model (Kaiser 1987) to the stochastic biasing model,
which is described in Appendix; we obtain $b^{(S)}_{\rm
\scriptscriptstyle obj} \lower.5ex\hbox{$\; \buildrel < \over \sim
\;$} b^{(R)}_{\rm \scriptscriptstyle obj}$ and $r^{(S)}_{\rm
\scriptscriptstyle obj}\simeq r^{(R)}_{\rm \scriptscriptstyle obj}$.
On small scales, however, the degree of the redshift-space distortion
depends sensitively on the nature of biased objects.  Note that the
redshift-space distortion even leads to $r^{(S)}_{\rm
\scriptscriptstyle halo}\gtsim 1$, but this is simply due to the
improper definition on small scales (see discussion in Section 3.4).
We also find that while $b^{(S)}_{\rm \scriptscriptstyle peak}$ is
significantly suppressed with respect to $b^{(R)}_{\rm
\scriptscriptstyle peak}$, the amplitude of $b^{(S)}_{\rm
\scriptscriptstyle halo}$ is comparable to that of $b^{(R)}_{\rm
\scriptscriptstyle halo}$.  For the latter aspect, this implies that
the pair-wise velocity dispersion, $\langle v_{12, \rm
\scriptscriptstyle obj}^2\rangle$, on small scales is dependent on
objects, and that $\langle v_{12, \rm \scriptscriptstyle
halo}^2\rangle \lower.5ex\hbox{$\; \buildrel < \over \sim \;$} \langle
v_{12, \rm \scriptscriptstyle mass}^2\rangle \lower.5ex\hbox{$\;
\buildrel < \over \sim \;$} \langle v_{12, \rm \scriptscriptstyle
peak}^2\rangle$. We directly compute the velocity dispersion and
indeed confirm that the above relation holds in our simulation
catalogues of peaks and halos.

\subsection{Time evolution}

Now we examine the time evolution of the biasing parameters.  For this
purpose, we consider the LCDM model and plot the biasing parameters as
a function of the redshift in Figure 6.  In order to probe the linear,
quasi-nonlinear and nonlinear scales of the gravitational clustering,
we select three separation lengths; $r=1$, $5$ and $23\, h^{-1}$Mpc
for peaks, and $r=7$, $13$ and $23\, h^{-1}$Mpc for halos. Again the
qualitative behavior is quite similar in the peak and halo models
regardless of the scale.  Toward lower redshifts, $r_{\rm
\scriptscriptstyle obj}$ increases and approaches unity while $b_{\rm
\scriptscriptstyle obj}$ decreases. We plot the number-conserving
model predictions (Tegmark, Peebles 1998) in Figure 6a using the
values of simulations at $z_2 =3.4$ for $b_{\rm \scriptscriptstyle
peak}(z_2)$ and $r_{\rm \scriptscriptstyle peak}(z_2)$ in equations
(\ref{eq:tp98bias}) and (\ref{eq:tp98coe}).  It turns out that the
time evolution of biasing in peak model is well described by the
number-conserving model even on fairly nonlinear scales.

Figure 6b plots the results in halo model. Since our halo catalogues
are selected by the mass, the number density of the selected halos
changes with redshift according to the Press-Schechter mass function.
Therefore the number-conserving model predictions do not match the
results. Rather we plot the bias formula by Jing (1998) which seem to
be in reasonable agreement with the evolution in simulations.
Incidentally we create a different halo catalogue which conserves the
number density by appropriately changing the selection mass threshold
with $z$, and make sure that the evolution in such halo catalogues
agrees well with the number-conserving model prediction.

\subsection{Comparison with biasing parameters in terms of the
 one-point statistics}

As noted earlier, the biasing parameters defined in equations
(\ref{eq:def_b}) and (\ref{eq:def_r}) are somewhat ambiguous on some
scales. This leads to unphysical results on small scales that the
cross-correlation factor $r_{\rm \scriptscriptstyle obj}$ does not lie
in the range $[-1,1]$ and that the biasing factor $b_{\rm
\scriptscriptstyle obj}$ may become imaginary.  Indeed, this behavior
$r_{\rm \scriptscriptstyle obj}\gtsim1$ shows up in the halo model
(Figure 5). In the light of this, it is useful and instructive to
relate our biasing parameters to those defined in terms of one-point
statistics which are wel-defined on all scales.

In general, the two-point statistics does not simply reduce to the
one-point statistics because of the difference of their smoothing
function.  Nevertheless in the case of the top-hat window function,
the variances of mass, objects and their cross-correlation can be
explicitly written in terms of the corresponding two-point correlation
functions as follows:
\begin{equation}
  \label{eq: def_sigma}
  \sigma_{ij}^2(R_{\rm \scriptscriptstyle TH}) = 
4\pi\int_{0}^{2R_{\rm \scriptscriptstyle TH}} dr\,\, r^2 \,\xi_{ij}(r)
  \,\, F(r;R_{\rm \scriptscriptstyle TH}), 
\end{equation}
where the subscripts $i$ and $j$ mean either ${\rm o}$ or ${\rm m}$. 
The filter function $F(r;R_{\rm \scriptscriptstyle TH})$ is given by 
\begin{equation}
  F(r;R_{\rm \scriptscriptstyle TH}) = 
\frac{3}{\pi}\,\frac{(r-2R_{\rm \scriptscriptstyle TH})^2
(r+4R_{\rm \scriptscriptstyle TH})}{(2R_{\rm \scriptscriptstyle
TH})^6} ,
\end{equation}
and satisfies the normalization condition :
\begin{equation}
  \label{eq:fnormalization}
4\pi \int_{0}^{2R_{\rm \scriptscriptstyle TH}} dr\,\, r^2 \,
 F(r;R_{\rm \scriptscriptstyle TH}) = 1 . 
\end{equation}

Using the above relation, one can in principle reproduce the biasing
and cross-correlation factors for one-point statistics in a
straightforward manner.  In practice, however, one has to take account
of the finite resolution of the numerical data carefully.  In particle
simulations, the smallest scales are supposed to be dominated by the
discreteness effect and the two-point correlation functions at $r=0$
are formally described by the sum of the Dirac delta function.  In
fact, this sensitively changes the amplitude of 
$\sigma_{ij}(R_{\rm \scriptscriptstyle TH})$ on small scales.  This
implies that one has to introduce a particular model of two-point
correlation functions at $r=0$.  For this purpose, we adopt the
following procedure.  We assume that the number density of each object
on smallest scales is simply described by the sum of (randomly
distributed) Dirac's delta function $n_{\rm\scriptscriptstyle obj}(x)=
\sum_{n}\delta_D(x-x_n)$. In this case, the Poisson process yields the
following offset to the intrinsic clustering term
$\sigma_{ii}^2(R_{\rm \scriptscriptstyle TH})$:
\begin{equation}
  \label{eq: offset}
  \sigma_{ii,{\rm true}}^2(R_{\rm \scriptscriptstyle TH}) 
= \sigma_{ii}^2(R_{\rm \scriptscriptstyle TH}) + 
  \sigma_{\rm shot}^2(R_{\rm \scriptscriptstyle TH}), 
\end{equation}
where the first term in the right-hand-side is calculated from
equation (\ref{eq: def_sigma}) using the numerical data of correlation
functions above the scales of the resolution. The second term is given
by
\begin{equation}
  \label{eq: sigma_shot}
\sigma_{\rm shot}^2(R_{\rm \scriptscriptstyle TH}) = 
\left[\frac{4\pi}{3}\,R_{\rm \scriptscriptstyle TH}^3 
\overline{n}_{\rm\scriptscriptstyle obj}
\right]^{-1}.
\end{equation}
in the case of the top-hat window function with
$\overline{n}_{\rm\scriptscriptstyle obj}$ being the mean number
density of objects.

Adopting equation (\ref{eq: offset}), we estimate the variances 
$\sigma_{ii,{\rm true}}$ and then compute the biasing parameters 
$b_{\rm\scriptscriptstyle obj}^{(1)}$ and 
$r_{\rm\scriptscriptstyle obj}^{(1)}$: 
\begin{equation}
  \label{eq: def_b1}
  b^{(1)}_{\rm\scriptscriptstyle obj}\equiv 
  \frac{\sigma_{\rm oo}}{\sigma_{\rm mm}}, 
\end{equation}
\begin{equation}
  \label{eq: def_r1}
  r^{(1)}_{\rm\scriptscriptstyle obj}\equiv 
  \frac{\sigma_{\rm om}^2}{\sigma_{\rm oo}\sigma_{\rm mm}}, 
\end{equation}
where $\sigma_{\rm om}= \sqrt{\langle\delta_{\rm\scriptscriptstyle
obj} \delta_{\rm\scriptscriptstyle mass}\rangle}$ is the one-point
cross-correlation.

In Figures 7 and 8, the biasing parameters for the one-point
statistics are shown as a function of smoothing radius $R_{\rm
\scriptscriptstyle TH}$.  On large scales, their behavior is similar
to that for the two-point statistics, and the theoretical prediction
is good agreement with the numerical results.  On the other hand, on
small scales, the point process dominates the clustering signal and
scale-dependence of the biasing and cross-correlation factor is quite
different from that of $b_{\rm\scriptscriptstyle obj}$ and
$r_{\rm\scriptscriptstyle obj}$ for both peaks and halos;
$r_{\rm\scriptscriptstyle obj}^{(1)}$ substantially deviates from
unity and $b_{\rm\scriptscriptstyle obj}^{(1)}$ becomes quite large as
the smoothing radius $R_{\rm\scriptscriptstyle TH}$ decreases. The
effect becomes more significant in the redshift-space and/or at higher
redshift, since the redshift-space distortion and redshift evolution
tend to weaken the intrinsic clustering amplitude.

The comparison of those biasing parameters between the one- and
two-point statistics suggests that our results are reliable on scales
larger than $\sim 1 h^{-1}$Mpc for peaks and $\sim 5h^{-1}$Mpc for
halos where the small-scale modeling does not seriously affect the
biasing parameters.

\section{Conclusions and discussion}

We have extensively examined the scale-dependence, time-evolution, and
redshift-space distortion of biasing in peak and halo models, with
particular attention to the nonlinear stochastic nature.  Especially,
we quantify the biasing properties defined in terms of the two-point
statistics and the results are compared with those in the one-point
statistics.  Our main results are summarized as follows:

(1) We have quantitatively demonstrated the degree of nonlinearity and
stochasticity of biasing in peak and halo models. In particular,
biasing is significantly nonlinear and stochastic in nonlinear regimes
of gravitational clustering, due to the gravitational evolution for peaks and 
volume exclusion for halos, which results in strong scale-dependence
of the biasing parameters on small scales ($\ltsim$ a few Mpc).

(2) On scales larger than $10h^{-1}$Mpc, however, a linear
deterministic and scale-independent biasing remains a reasonable
approximation in peak and halo models.  While the biasing parameters
in this linear regime evolve as a redshift, their evolution is
consistent with the existing theoretical model predictions.

(3) Redshift-space distortion of the biasing on small scales is quite
different in peak and halo models because the pair-wise velocity
dispersions are quite sensitive to those objects.  This difference
affects the stochasticity of biasing in redshift space.  The linear
redshift-space distortion, on the other hand, is very similar and
explained by extending the model of Kaiser (1987) to the stochastic
biasing case (see Appendix).

The biasing models considered here are rather idealistic for the
observed luminous objects which form through the complicated processes
including the gas dynamics and the radiative transfer.  Nevertheless
the peak and halo models contain the most important aspect of the
formation of astrophysical objects, the nonlinear gravitational growth
and the subsequent merging, in the scenario of hierarchical galaxy
formation.  This is why our results are qualitatively similar to the
recent numerical work of C\'olin et al. (1999) and Somerville et
al. (1999), and also to predictions of an analytical model of Taruya
\& Suto (2000).  Moreover cosmological simulations of galaxy formation
by Yoshikawa et al. (2000) indicate that the one-to-one correspondence
between halos and galaxies may be a reasonably good approximation on
large scales and the nonlinear clustering of peaks traces the galaxy
distribution fairly well even on small scales.

The validity of linear deterministic and scale-independent relation on
large scales may be a natural consequence of the hierarchical
clustering scenario (e.g., Matsubara 1999). Such scale-independence
probably comes from the fact that the statistical feature of
large-scale clustering of peaks and halos is simply characterized by a
single parameter; the mass threshold $M$ for halos and the density
height $\nu$ for peaks.  Our important result is that the relation
$\delta_{\rm \scriptscriptstyle obj}\simeq b_{\rm \scriptscriptstyle
obj} \delta_{\rm \scriptscriptstyle mass}$ still holds in quasi-linear
regime.  If this is the case, the interpretation of clustering of
realistic objects is straightforward.  On the other hand, the
scale-dependence of the biasing becomes crucial in the nonlinear
regime of density fluctuations.  These should be kept in mind in
interpreting the future redshift surveys of galaxies.

\vspace{1pc}\par

We thank Ue-Li Pen, for pointing out the importance of the relation
between the one- and two-point statistics (eq.[\ref{eq: def_sigma}]),
which improves our analysis and leads to the discussion in Section
3.4.  Y. P. J. and A.T. gratefully acknowledge support from a JSPS
(Japan Society for the Promotion of Science) fellowship.  Numerical
computations were carried out on VPP300/16R and VX/4R at ADAC (the
Astronomical Data Analysis Center) of the National Astronomical
Observatory, Japan, as well as at RESCEU (Research Center for the
Early Universe, University of Tokyo) and KEK (High Energy Accelerator
Research Organization, Japan). This research was supported in part by
the Grant-in-Aid by the Ministry of Education, Science, Sports and
Culture of Japan (07CE2002, 12640231) to RESCEU, and by the
Supercomputer Project (No.99-52, No.00-63) of KEK.

\section*{Appendix.\ Linear stochastic biasing in redshift-space}

\setcounter{equation}{0}
\renewcommand{\theequation}{A\arabic{equation}}

This appendix describes the generalization of the the linear
redshift-space distortion model of Kaiser (1987) in the context of the
stochastic biasing.  Although we consider the biasing for the 
one-point statistics, the results are also applicable to the 
biasing for two-point statistics as long as the correlation functions 
are positive.

First recall that the density contrast of dark
matter and objects in k-space is given by
\begin{eqnarray}
\label{eq: delta_m_red}
    \delta_{\rm \scriptscriptstyle mass}^{\rm(S)}({\bf k}) 
&=& \delta_{\rm \scriptscriptstyle mass}^{\rm(R)}({\bf k})
+f\mu^2\delta_{\rm \scriptscriptstyle mass}^{\rm(R)}({\bf k}) , \\
\label{eq: delta_o_red}
    \delta_{\rm \scriptscriptstyle obj}^{\rm(S)}({\bf k}) 
&=& \delta_{\rm \scriptscriptstyle obj}^{\rm(R)}({\bf k})
+f\mu^2\delta_{\rm \scriptscriptstyle mass}^{\rm(R)}({\bf k}) ,
\end{eqnarray}
where $\mu=k_{\scriptscriptstyle\parallel}/k$ is the direction cosine
in k-space ($k_{\scriptscriptstyle\parallel}$ is the comoving wave
number parallel to the line of sight), and
\begin{eqnarray} 
\label{eq:fz}
   f &\equiv& \frac{d\ln D(z)}{d\ln a} \simeq
   \Omega(z)^{0.6} + {\lambda(z) \over 70} 
\left(1+ {\Omega(z) \over 2}\right), \\
   \Omega(z) &=& \left[{H_0 \over H(z)}\right]^2 (1+z)^3\, \Omega_0 , \\
   \lambda(z) &=& \left[{H_0 \over H(z)}\right]^2 \lambda_0 .
\end{eqnarray}
Equations (\ref{eq: delta_m_red}) and (\ref{eq: delta_o_red}), relate
the auto- and cross-correlation functions in redshift space to those
in real space:
\begin{eqnarray}
\label{eq: mass_mass}
 \langle \delta_{\rm \scriptscriptstyle mass}^{\rm(S)}
\delta_{\rm \scriptscriptstyle mass}^{\rm(S)} \rangle (\mu)
  &=& \left(1+2f\mu^2+f^2\mu^4\right) 
\langle \delta_{\rm \scriptscriptstyle mass}^{\rm(R)}
  \delta_{\rm \scriptscriptstyle mass}^{\rm(R)} \rangle, 
\\
\label{eq: obj_obj}
 \langle \delta_{\rm \scriptscriptstyle obj}^{\rm(S)}
\delta_{\rm \scriptscriptstyle obj}^{\rm(S)} \rangle (\mu)
  &=& \left(1+2\beta r_{\rm \scriptscriptstyle obj}^{\rm(R)}\mu^2
+\beta^2\mu^4\right) \langle 
  \delta_{\rm \scriptscriptstyle obj}^{\rm(R)}
\delta_{\rm \scriptscriptstyle obj}^{\rm(R)} \rangle, 
\\
\label{eq: mass_obj}
 \langle \delta_{\rm \scriptscriptstyle mass}^{\rm(S)}
\delta_{\rm \scriptscriptstyle obj}^{\rm(S)} \rangle (\mu)
  &=& \left[b_{\rm \scriptscriptstyle obj}^{\rm(R)}
r_{\rm \scriptscriptstyle obj}^{\rm(R)}
+f\mu^2\left(1+b_{\rm \scriptscriptstyle obj}
  ^{\rm(R)}r_{\rm \scriptscriptstyle obj}^{\rm(R)}\right)+
  f^2\mu^4\right] \langle \delta_{\rm \scriptscriptstyle mass}^{\rm(R)}
\delta_{\rm \scriptscriptstyle mass}^{\rm(R)} \rangle ,
\end{eqnarray}
where the brackets indicate the ensemble average, 
$\beta \equiv
f/b_{\rm \scriptscriptstyle obj}^{\rm(R)}$ and  
the quantities $b_{\rm \scriptscriptstyle obj}^{\rm(R)}$ and 
$r_{\rm \scriptscriptstyle obj}^{\rm(R)}$ denote the 
biasing and cross-correlation factors in real space:
\begin{eqnarray}
    b_{\rm \scriptscriptstyle obj}^{\rm(R)} 
&\equiv&
  \sqrt{\frac{\langle\delta_{\rm \scriptscriptstyle obj}^{\rm(R)}
\delta_{\rm \scriptscriptstyle obj}^{\rm(R)}\rangle}
  {\langle \delta_{\rm \scriptscriptstyle mass}^{\rm(R)}
\delta_{\rm \scriptscriptstyle mass}^{\rm(R)}\rangle}}, \\
    r_{\rm \scriptscriptstyle obj}^{\rm(R)} 
&\equiv &
     \frac{\langle \delta_{\rm \scriptscriptstyle mass}^{\rm(R)}
\delta_{\rm \scriptscriptstyle obj}^{\rm(R)} \rangle}
     {\sqrt{\langle \delta_{\rm \scriptscriptstyle mass}^{\rm(R)}
\delta_{\rm \scriptscriptstyle mass}^{\rm(R)} \rangle
     \langle \delta_{\rm \scriptscriptstyle obj}^{\rm(R)}
\delta_{\rm \scriptscriptstyle obj}^{\rm(R)} \rangle}} .
\end{eqnarray}
The biasing and cross-correlation factors in redshift space are
defined similarly using equations (\ref{eq: mass_mass})-(\ref{eq:
  mass_obj}).  After integrating over $\mu$, we find
\begin{eqnarray}
\label{eq: bias_red}
 b^{\rm(S)} &\equiv&
  \sqrt{\frac{\langle\delta_{\rm \scriptscriptstyle obj}^{\rm(S)}
\delta_{\rm \scriptscriptstyle obj}^{\rm(S)}\rangle}
  {\langle \delta_{\rm \scriptscriptstyle mass}^{\rm(S)}
\delta_{\rm \scriptscriptstyle mass}^{\rm(S)}\rangle}}
 = b_{\rm \scriptscriptstyle obj}^{\rm(R)}
\sqrt{\frac{(1+\frac{2}{3}\beta r_{\rm \scriptscriptstyle obj}^{\rm(R)}+
  \frac15\beta^2)}{(1+\frac{2}{3}f+\frac15f^2)}} , \\
\label{eq: cross_red}
    r_{\rm \scriptscriptstyle obj}^{\rm(S)} &\equiv&
 \frac{\langle \delta_{\rm \scriptscriptstyle mass}^{\rm(S)}
     \delta_{\rm \scriptscriptstyle obj}^{\rm(S)} \rangle} 
{\sqrt{\langle \delta_{\rm \scriptscriptstyle mass}
     ^{\rm(S)}\delta_{\rm \scriptscriptstyle mass}^{\rm(S)} \rangle 
     \langle\delta_{\rm \scriptscriptstyle obj}^{\rm(S)}
\delta_{\rm \scriptscriptstyle obj}^{\rm(S)} \rangle}} 
 = \frac{r_{\rm \scriptscriptstyle obj}^{\rm(R)}
+\frac13\beta(1+b_{\rm \scriptscriptstyle obj}^{\rm(R)}
  r_{\rm \scriptscriptstyle obj}^{\rm(R)})+\frac15f\beta}
  {\sqrt{(1+\frac{2}{3}f+\frac15f^2)
(1+\frac{2}{3}\beta r_{\rm \scriptscriptstyle obj}
  ^{\rm(R)}+\frac15\beta^2)}} .
\end{eqnarray}

In the deterministic case $r^{(R)}_{\rm \scriptscriptstyle obj}=1$, 
one can show that
\begin{eqnarray}
\label{eq:bsbr}
 b^{(S)}_{\rm \scriptscriptstyle obj} &\leq&
 b^{(R)}_{\rm \scriptscriptstyle obj}, \\
\label{eq:rsrr}
 r^{(S)}_{\rm \scriptscriptstyle obj} &\leq& 
r^{(R)}_{\rm \scriptscriptstyle obj} = 1 , 
\end{eqnarray}
from equations (\ref{eq: bias_red}) and (\ref{eq: cross_red}).  The
inequality (\ref{eq:bsbr}) is understood by the squashing effect which
is more appreciable for $\delta^{(S)}_{\rm \scriptscriptstyle mass}$
if $\delta^{(R)}_{\rm\scriptscriptstyle obj} >
\delta^{(R)}_{\rm\scriptscriptstyle mass}$ (see eqs.[\ref{eq:
delta_m_red}] and [\ref{eq: delta_o_red}]).  The inequality
(\ref{eq:rsrr}) merely reflects the fact that the random peculiar
velocity generates an additional scatter around the mean biasing
relation in redshift space. 

In a more general case, the biasing is not deterministic i.e.,
$r^{(R)}_{\rm \scriptscriptstyle obj} <1$.  Even then the inequality
(\ref{eq:bsbr}) still holds as long as $b^{(R)}_{\rm
\scriptscriptstyle obj}>1$ and $r^{(R)}_{\rm \scriptscriptstyle
obj}\leq1$. The inequality (\ref{eq:rsrr}), on the other hand, no
longer holds, and in fact, the opposite inequality seems to be a
fairly general result; Figure 9 plots $r^{(S)}_{\rm \scriptscriptstyle
obj}-r^{(R)}_{\rm \scriptscriptstyle obj}$ as a function of
$r^{(R)}_{\rm \scriptscriptstyle obj}$ for different values of $f$;
$f=0.3$ (solid lines), $f=0.6$ (dashed lines), and $0.9$ (dotted
lines) with $b^{(R)}_{\rm \scriptscriptstyle obj}=3.0$ ({\it
left-panel}) and $b^{(R)}_{\rm \scriptscriptstyle obj}=5.0$ ({\it
right-panel}). Except for a small region around $r^{(R)}_{\rm
\scriptscriptstyle obj}\simeq 1$, $r^{(S)}_{\rm \scriptscriptstyle
obj}$ is generally {\it larger} than $r^{(R)}_{\rm \scriptscriptstyle
obj}$. It seems that the stochasticity is effectively reduced due to
the redshift-space distortion as increasing the stochasticity in real
space. Moreover this tendency becomes stronger when the factor $f$
becomes large, or equivalently, the density parameter $\Omega_0$
approaches unity.

Thus in linear theory of redshift-space distortion, we conclude fairly
generally that $b^{(S)}_{\rm \scriptscriptstyle obj} \leq b^{(R)}_{\rm
\scriptscriptstyle obj}$ and $r^{(S)}_{\rm \scriptscriptstyle obj}
\lower.5ex\hbox{$\; \buildrel > \over \sim \;$} r^{(R)}_{\rm
\scriptscriptstyle obj}$ as long as a certain degree of the
stochascity exists in real space (i.e., $r^{(R)}_{\rm
\scriptscriptstyle obj}< 1$) as Figure 9 exhibits.

\bigskip
\bigskip

\section*{References}

\re
  Bardeen J.M., Bond J.R., Kaiser N., Szalay, A.S.\ 1986, ApJ\ 304, 15
\re
  Blanton M., Cen R.Y., Ostriker J.P., Strauss M.A.\ 1999, ApJ\ 522, 590
\re
  Blanton M., Cen R.Y., Ostriker J.P., Strauss M.A., Tegmark M.\ 2000,
  ApJ\ 531, 1
\re
  C\'olin P., Klypin A.A., Kravtsov A.V., Khokhlov A.\  1999, ApJ\ 523, 32  
\re
  Dekel A., Lahav O.\ 1999, ApJ\ 520, 24
\re
  Fry J.N.\ 1996, ApJL\ 461, 65
\re
  Hamilton A.J.S. \ 1998, in Proceedings of Ringberg Workshop on
  Large-Scale Structure, Hamilton, D.(ed.), Kluwer Academic, Dordrecht, p.185  
\re
  Jing Y.P.\ 1998, ApJL\ 503, 9
\re
  Jing Y.P., Suto Y.\ 1998,ApJL\ 494, 5
\re
  Kaiser N.\ 1984, ApJL\ 284, 9
\re
  Kaiser N.\ 1987, MNRAS\ 227, 1
\re
  Kitayama T., Suto Y.\ 1997, ApJ\ 490, 557
\re
  Matsubara T.\ 1999, ApJ\ 525, 543
\re
  Mo H.J., White, S.D M.\ 1996, MNRAS\ 282, 347
\re
  Mo H.J., Jing Y.P., White, S.D.M.\ 1997, MNRAS\ 284, 189
\re
  Peebles P.J.E. \ 1980, The Large Scale Structure of the Universe, 
  Princeton University Press
\re
  Pen U.L. \ 1998, ApJ\ 504, 601 
\re
  Press W.H., Schechter P.\ 1974, ApJ\ 187, 425
\re
  Somerville R.S., Lemson G., Sigad Y., Dekel A., Kauffmann G., White S.D.M.\ 1999, preprint (astro-ph/9912073)
\re
  Taruya A.\ 2000, ApJ\ 537, 37
\re
  Taruya A., Koyama K, Soda J.\ 1999,ApJ\ 510, 541
\re
  Taruya A., Soda J.\ 1999, ApJ\ 522, 46
\re
  Taruya A., Suto Y.\ 2000, ApJ\ 542, 559
\re
  Tegmark M., Peebles P.J.E.\ 1998, ApJL\ 500, 79
\re
  Tegmark M., Bromley B. \ 1999, ApJL\ 518, 69 
\re
  Yoshikawa K., Taruya A., Jing Y.P., Suto Y.\ 2000, ApJ, submitted 

\clearpage

\begin{table*}[t]
\begin{center}
Table~1.\hspace{4pt}Simulation model parameters.\\
\end{center}
\vspace{6pt}
\begin{tabular*}{\textwidth}{@{\hspace{\tabcolsep}\extracolsep{\fill}}lcccccccc}
\hline\hline\\[-6pt]
Model & $\Omega_0$ &  $\lambda_0$  
&  $\Gamma^{\dagger}$ & $\sigma_8$  & $N$ & $m_{\rm p}^\ddagger [h^{-1}{\rm M}_\odot]$ & realizations \\[4pt] 
\hline\\[-6pt]
SCDM (Standard CDM) & 1.0 & 0.0 & 0.5  & 0.6 & $256^3$ & $4.5\times10^{11}$ & 3  \\
LCDM (Lambda CDM)   & 0.3 & 0.7 & 0.21 & 1.0 & $256^3$ & $1.3\times10^{11}$ & 3  \\
OCDM (Open CDM)     & 0.3 & 0.0 & 0.25 & 1.0 & $256^3$ & $1.3\times10^{11}$ & 3  \\
\hline 
\end{tabular*}
\vspace{6pt}\par\noindent
$^{\dagger}$ the shape parameter of the power spectrum. 
\par\noindent
$^{\ddagger}$ mass of a single dark matter particle.
\end{table*}

\begin{table*}[t]
\begin{center}
Table~2.\hspace{4pt}The peak catalogue.
\end{center}
\begin{tabular*}{\textwidth}{@{\hspace{\tabcolsep}\extracolsep{\fill}}lcccccc}
\hline\hline\\[-6pt]
       & \multicolumn{2}{c}{SCDM} & \multicolumn{2}{c}{LCDM} & \multicolumn{2}{c}{OCDM} \\  
$\nu_{\rm th}$ & $N^{\dagger}$ & $b_{\rm \scriptscriptstyle peak, eff}^{\ddagger}$ 
 & $N$ & $b_{\rm \scriptscriptstyle peak, eff}$ & $N$ & $b_{\rm \scriptscriptstyle peak, eff}$ \\
\hline\\[-6pt]
1.0 & $6.2\times10^{5}$ & 1.2 (1.8) & $5.3\times10^{5}$ & 1.2 (1.5) &
 $5.5\times10^{5}$ & 1.2 (1.4) \\
2.0 & $2.6\times10^{5}$ & 1.6 (2.8) & $2.2\times10^{5}$ & 1.5 (2.2) & 
 $2.3\times10^{5}$ & 1.5 (1.9) \\
3.0 & $5.0\times10^{4}$ & 2.0 (4.1) & $4.5\times10^{4}$ & 1.8 (3.1) &  
 $4.6\times10^{4}$ & 1.8 (2.6) \\
\hline
\end{tabular*}
\vspace{6pt}\par\noindent
$^{\dagger}$ average total number of peaks among three realizations.
\par\noindent
$^{\ddagger}$ effective biasing factor of peaks at $z=0.0$ ($z=2.2$). 
\end{table*}

\begin{table*}[t]
\begin{center}
Table~3.\hspace{4pt}The halo catalogue.
\end{center}
\begin{tabular*}{\textwidth}{@{\hspace{\tabcolsep}\extracolsep{\fill}}lcccccc}
\hline\hline\\[-6pt]
       & \multicolumn{2}{c}{SCDM} & \multicolumn{2}{c}{LCDM} & \multicolumn{2}{c}{OCDM} \\  
$M_{\rm th}[h^{-1}{\rm M}_\odot]$ & $N^{\dagger}$ & $b_{\rm \scriptscriptstyle halo,eff}^{\ddagger}$ & $N$ & $b_{\rm \scriptscriptstyle halo,eff}$ & $N$ & $b_{\rm \scriptscriptstyle halo,eff}$ \\
\hline\\[-6pt]
$2\times10^{12}$ & ----------- \,(-----------) & 1.1 (5.2) &
 $7.3\times10^{4}$ $(3.9\times10^{4})$ & 0.9 (2.6) & $8.2\times10^{4}$ $(6.6\times10^{4})$& 0.9 (1.8)\\
$5\times10^{12}$ & $1.2\times10^{5}$ $(1.4\times10^{4})$ & 1.1 (5.2) &
 $3.0\times10^{4}$ $(1.1\times10^{4})$ & 1.0 (3.3) & $3.4\times10^{4}$ $(2.1\times10^{4})$ & 1.0 (2.3)\\
$1\times10^{13}$ & $5.7\times10^{4}$ $(2.3\times10^{3})$ & 1.3 (6.7) &
 $1.5\times10^{4}$ $(3.3\times10^{3})$ & 1.2 (4.1) &  $1.7\times10^{4}$ $(8.3\times10^{4})$ & 1.2 (2.8)\\
\hline
\end{tabular*}
\vspace{6pt}\par\noindent
$^{\dagger}$ average total number of halos among three realizations at
 $z=0.0$ ($z=2.2$).
\par\noindent
$^{\ddagger}$ effective biasing factor of halos at $z=0.0$ ($z=2.2$).
\end{table*}

\clearpage

\begin{figure}[htb]
\begin{center}
   \leavevmode\epsfxsize=0.65\textwidth \epsfbox{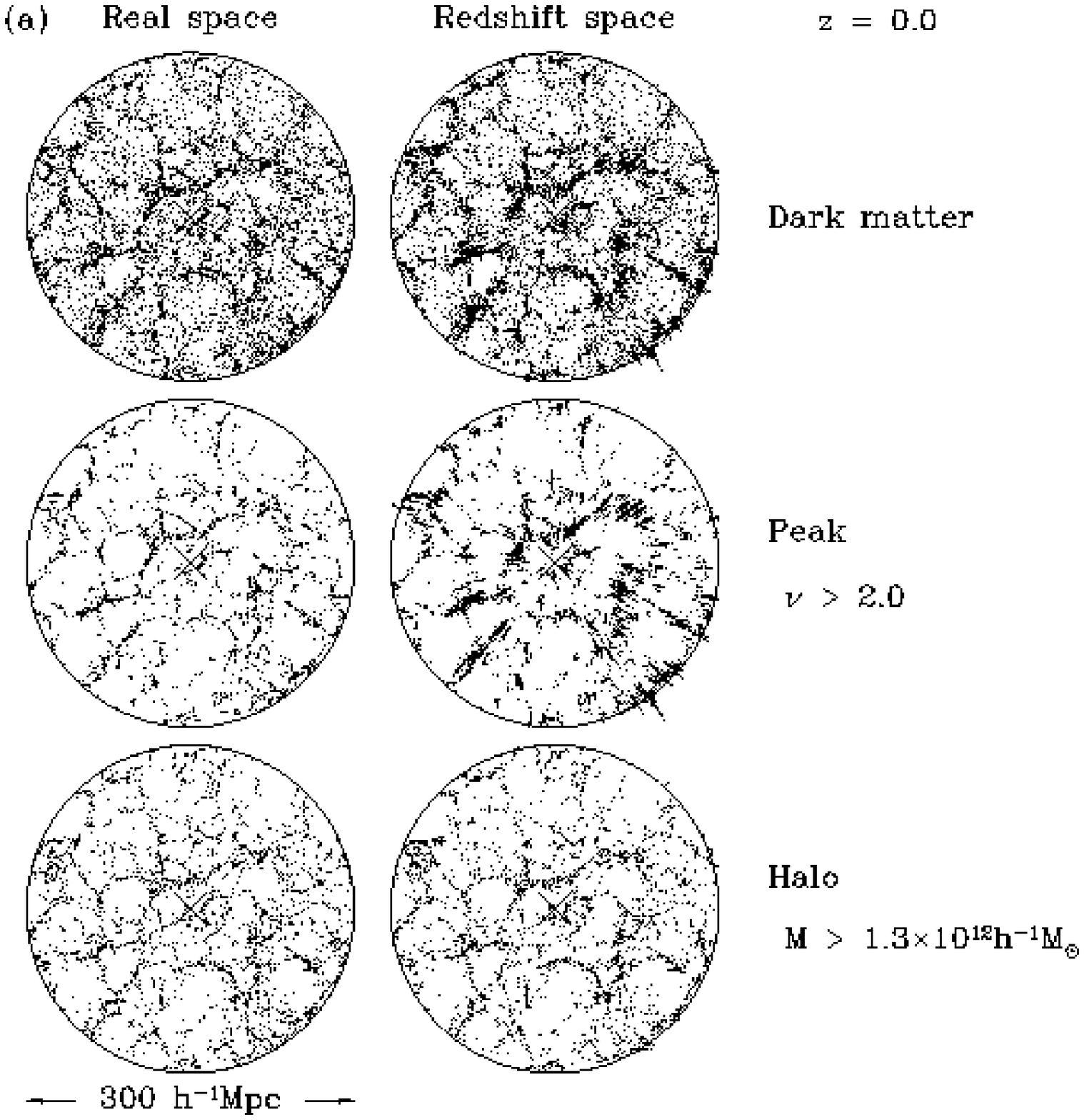} 
\end{center}
\begin{center}
   \leavevmode\epsfxsize=0.65\textwidth \epsfbox{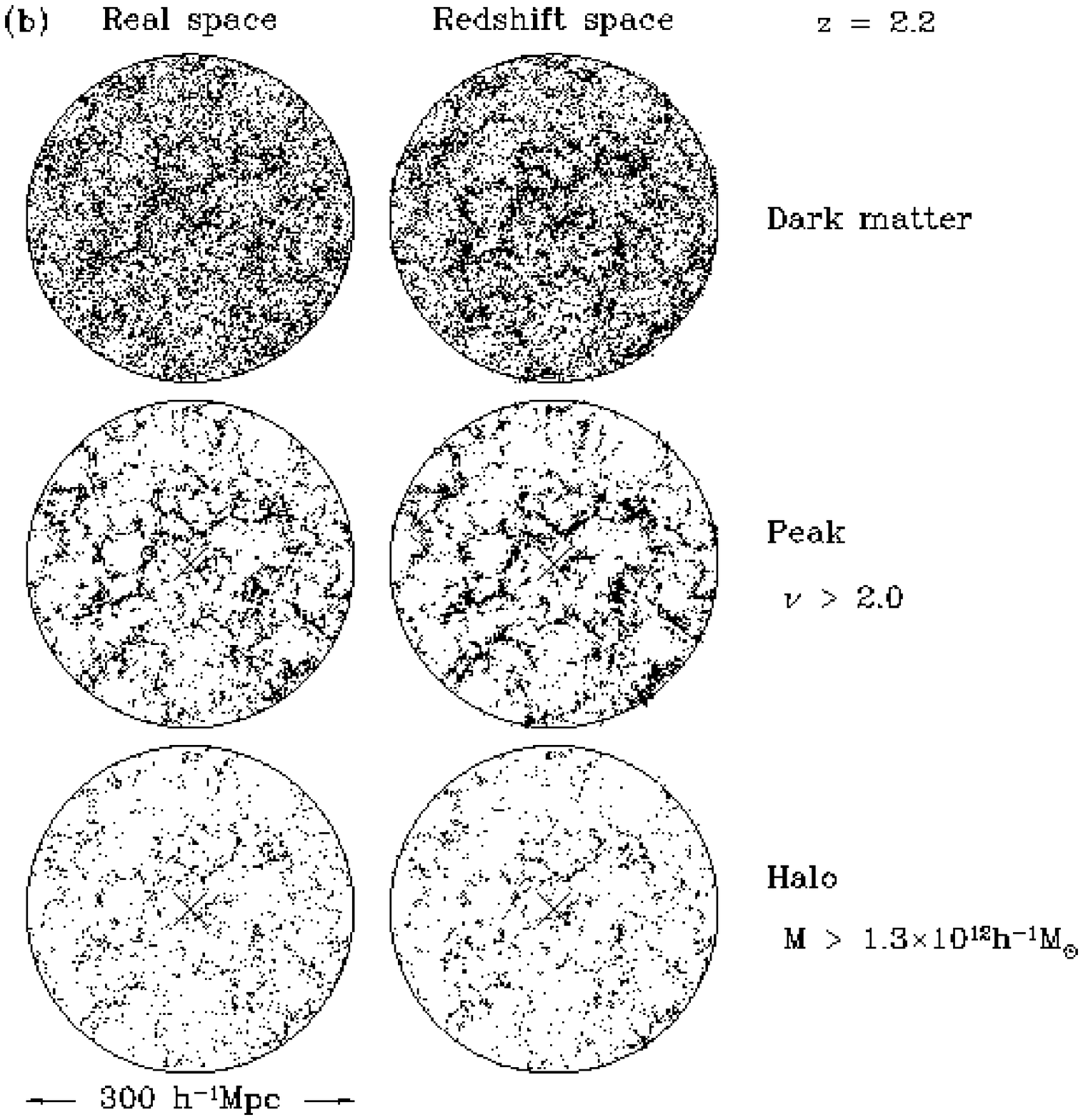}
\caption{Top-view of distribution of objects in real (left panels) and
  redshift (right panels) spaces around the fiducial observer at the
  center; dark matter particles (top panels), peaks with $\nu>2$
  (middle panels) and halos with $M>1.3\times10^{12}{\rm M}_\odot$
  (bottom panels) in LCDM model. The thickness of those slices is
  15$h^{-1}$Mpc; (a) $z=0$, (b) $z=2.2$.}
\end{center}
\end{figure}

\begin{figure}[ht]
\begin{center}
   \leavevmode\epsfxsize=0.48\textwidth \epsfbox{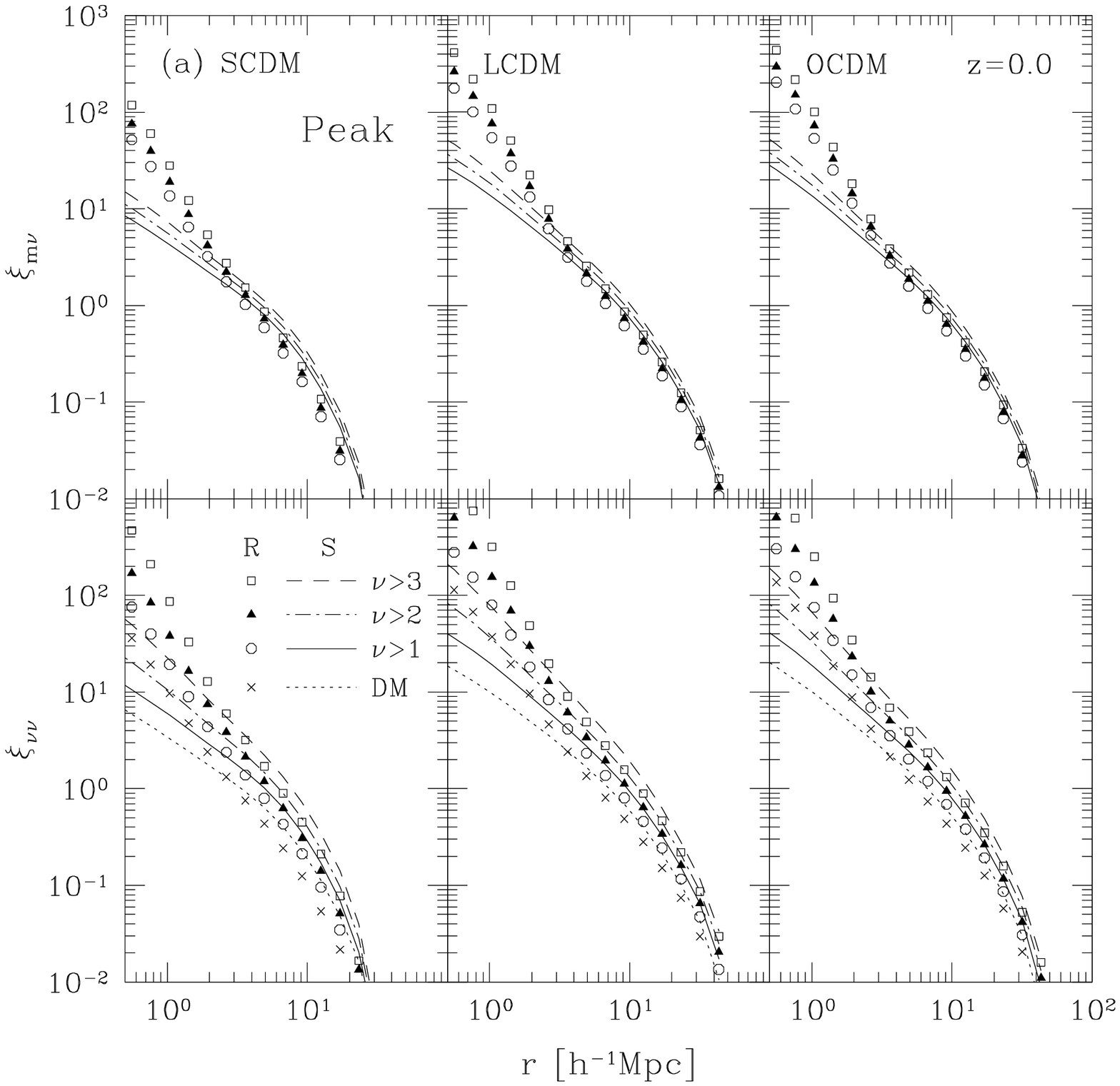}
   \leavevmode\epsfxsize=0.48\textwidth \epsfbox{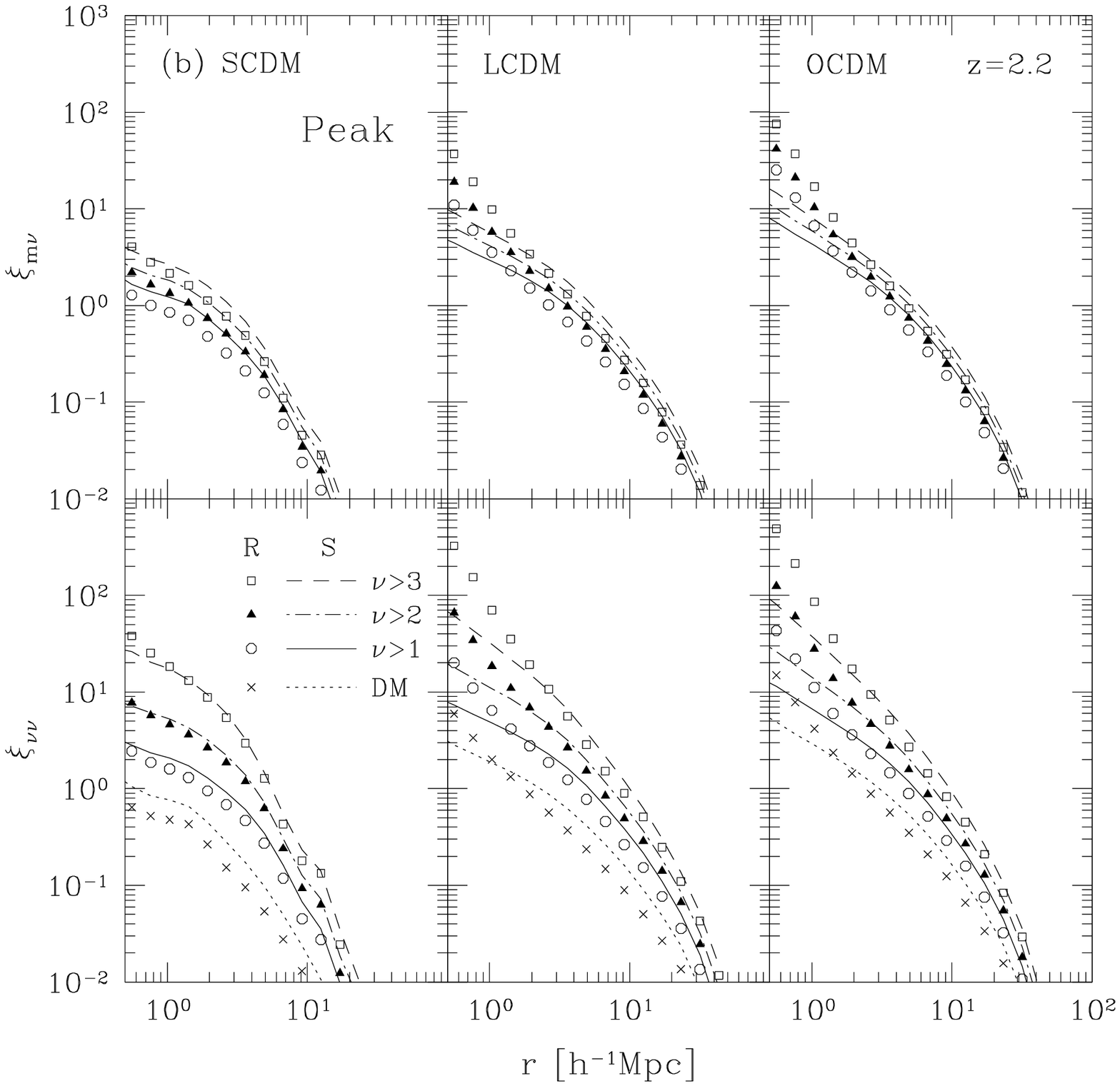}
\caption{Auto- and cross-correlation functions of dark matter and peaks
  in SCDM (left panels), LCDM (middle panels) and OCDM (right panels).
  Different symbols indicate the results in real space (open squares
  for $\nu>3$, filled triangles for $\nu>2$, open circles for $\nu>1$,
  and crosses for dark matter), while different curves indicate those in
  redshift space (dashed for $\nu>3$, dot--dashed for $\nu>2$,
solid for $\nu>1$, and dotted for dark matter).
(a) $z=0$, (b) $z=2.2$.
}
\end{center}
\end{figure}

\begin{figure}[ht]
\begin{center}
   \leavevmode\epsfxsize=0.48\textwidth \epsfbox{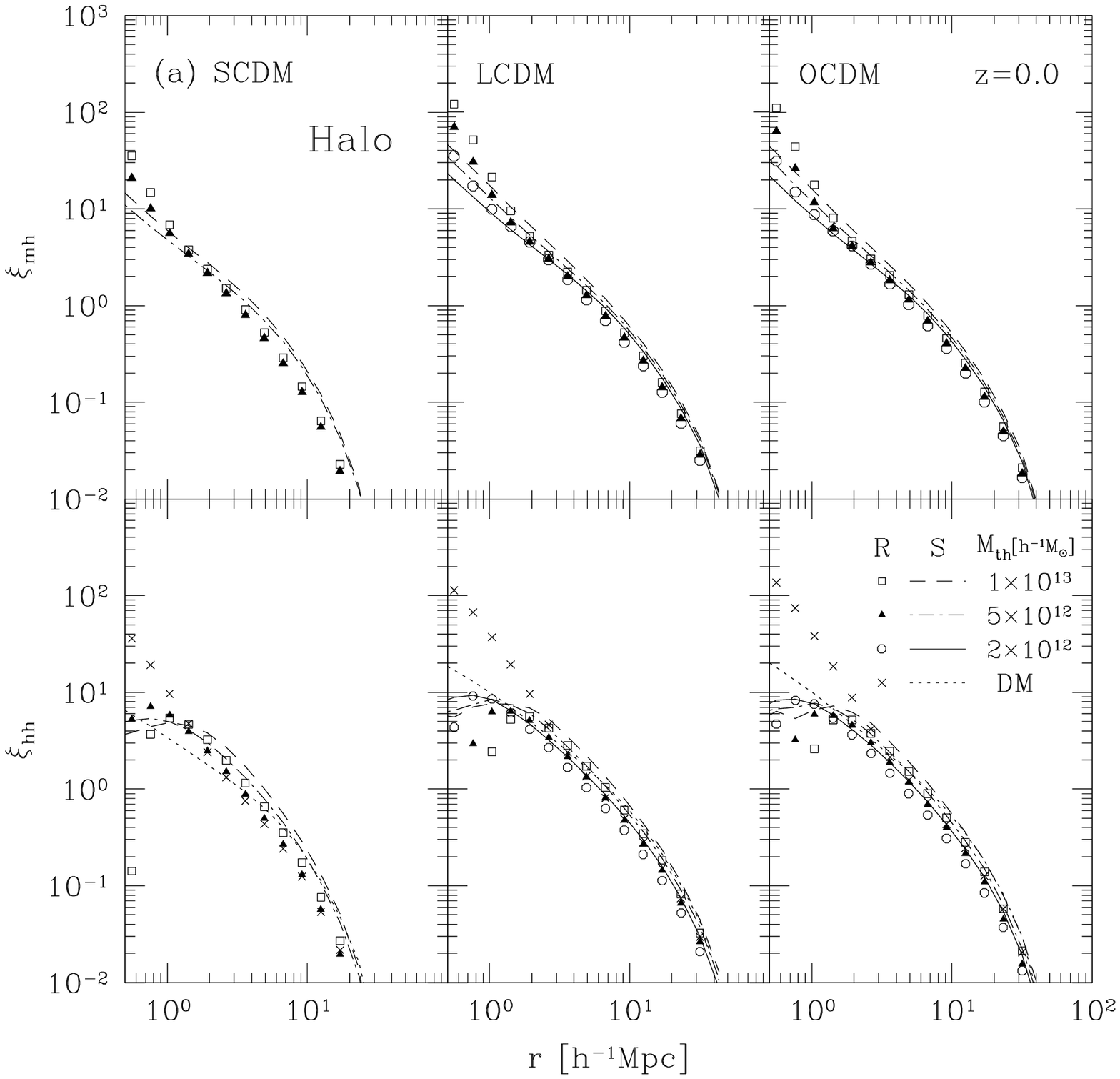}
   \leavevmode\epsfxsize=0.48\textwidth \epsfbox{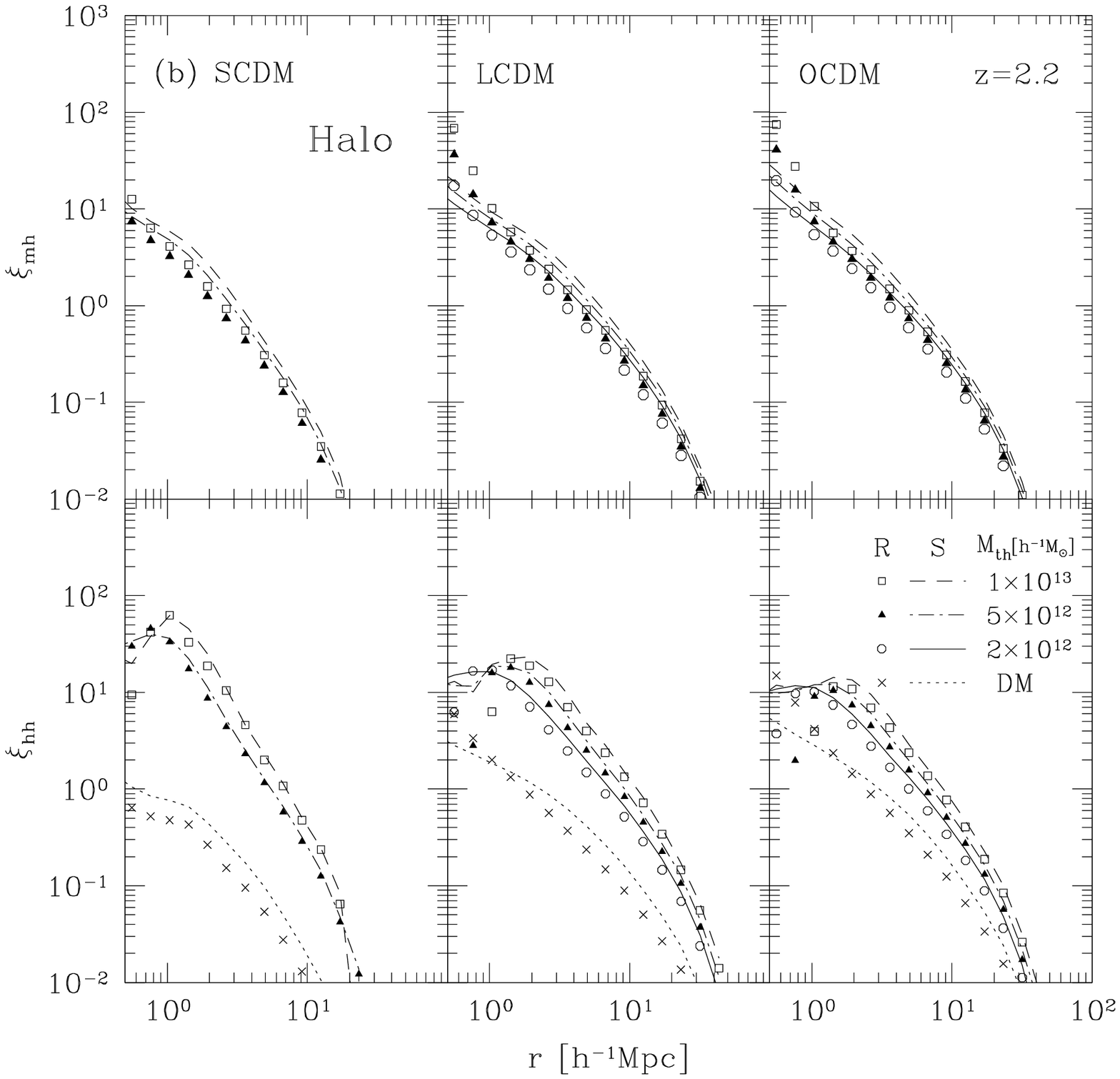}
\caption{Same as Fig. 2 but for halo model;
  open squares and dashed lines for $M > 10^{13}h^{-1}M_\odot$, filled
  triangles and dot--dashed lines for $M > 5\times
  10^{12}h^{-1}M_\odot$, open circles and solid lines for $M > 2\times
  10^{12}h^{-1}M_\odot$, and crosses and dotted lines for dark matter. 
  For SCDM model, we only plot the correlation functions with 
  $M_{\rm th}=5\times 10^{12},\,\,10^{13}h^{-1}M_\odot$. 
(a) $z=0$, (b) $z=2.2$.}
\end{center}
\end{figure}

\begin{figure}[ht]
\begin{center}
   \leavevmode\epsfxsize=0.48\textwidth \epsfbox{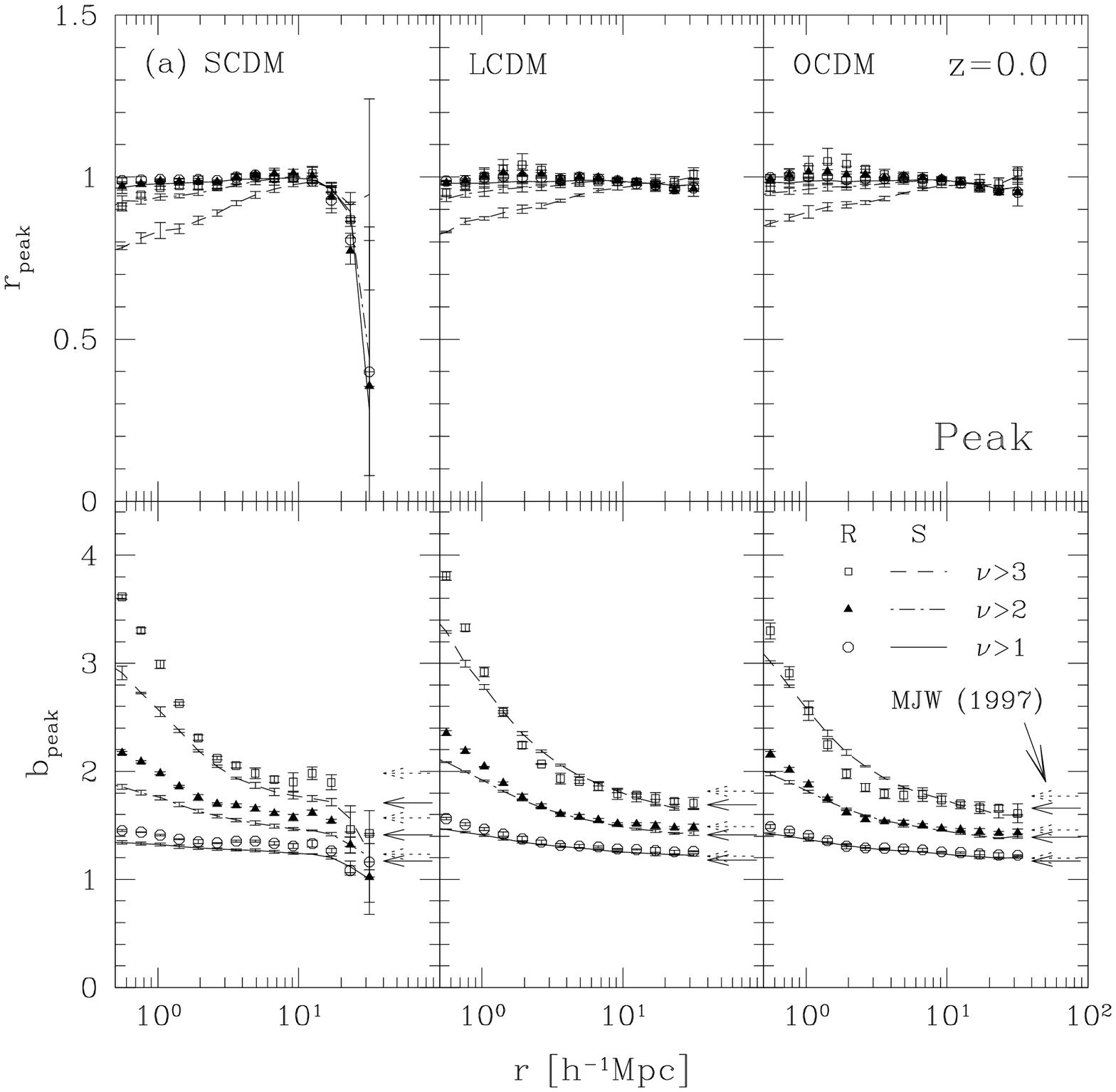}
   \leavevmode\epsfxsize=0.48\textwidth \epsfbox{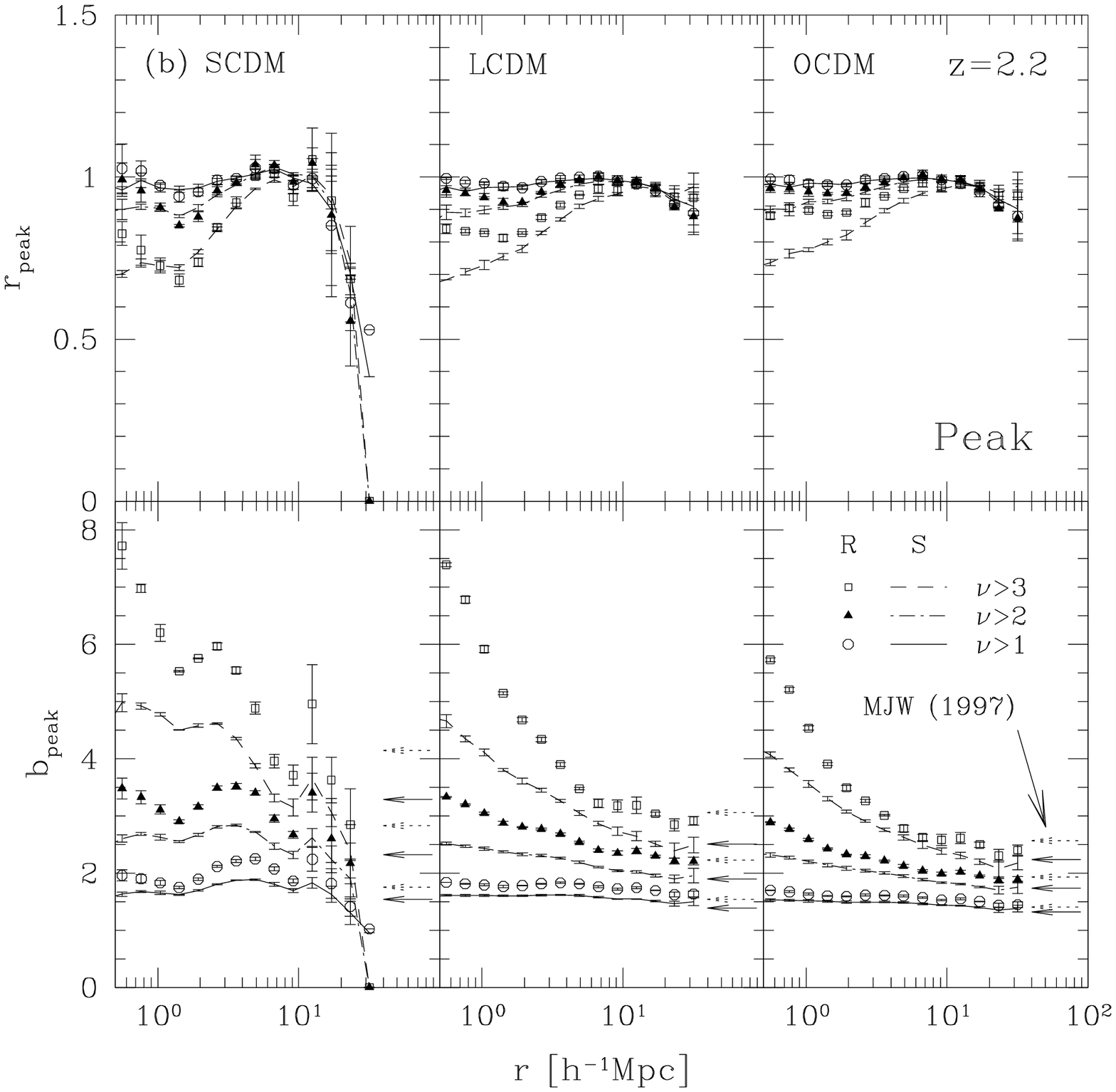}
\caption{Scale-dependence of the biasing and cross-correlation
  factors for two-point statistics of peak model. 
  Different symbols and lines have the same
  meaning as in Fig.2.  The arrows in the bottom panels indicate the
  model predictions of Mo, Jing, \& White (1997) in real (dotted) and
  redshift (solid) spaces.
(a) $z=0$, (b) $z=2.2$. }
\end{center}
\end{figure}

\begin{figure}[ht]
\begin{center}
   \leavevmode\epsfxsize=0.48\textwidth \epsfbox{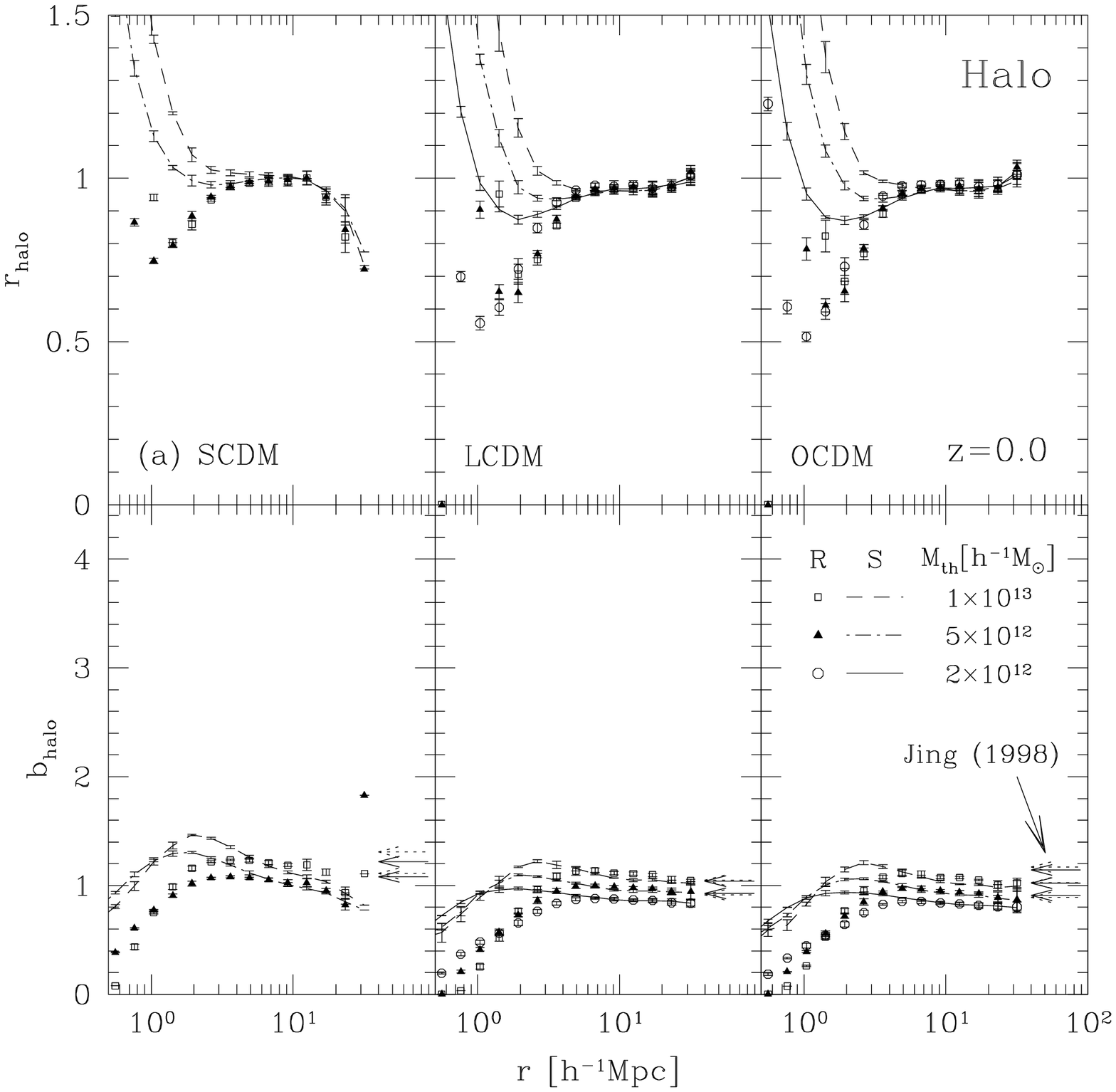}
   \leavevmode\epsfxsize=0.48\textwidth \epsfbox{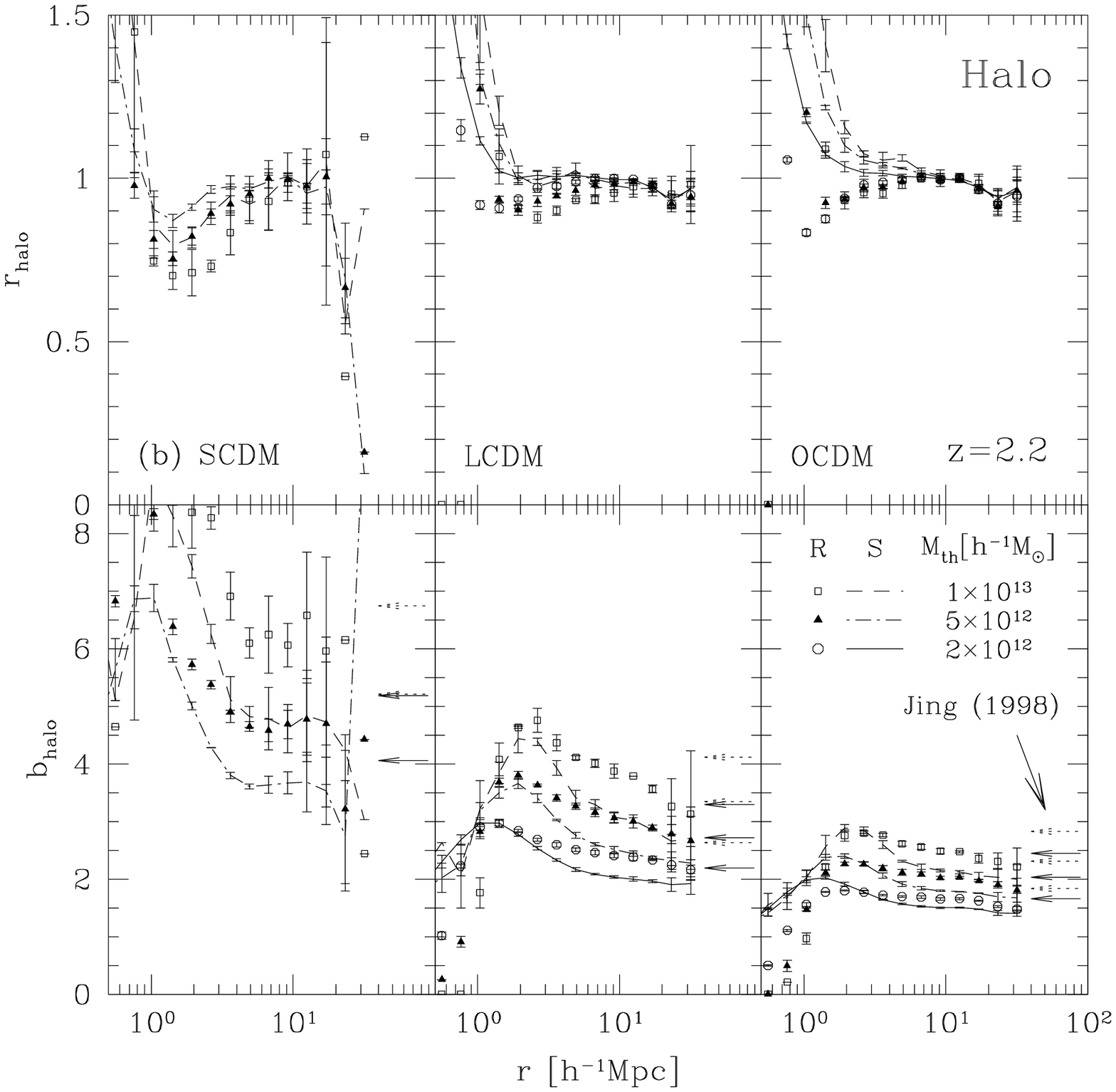}
\caption{Same as Fig. 4, but for halo model.  
  Different symbols and lines have the same
  meaning as in Fig.3. The arrows in the bottom panels indicate the
  model predictions of Jing (1998) in real (dotted) and redshift
  (solid) spaces.  (a) $z=0$, (b) $z=2.2$. }
\end{center}
\end{figure}

\clearpage
\begin{figure}[hb]
\begin{center}
   \leavevmode\epsfxsize=0.48\textwidth \epsfbox{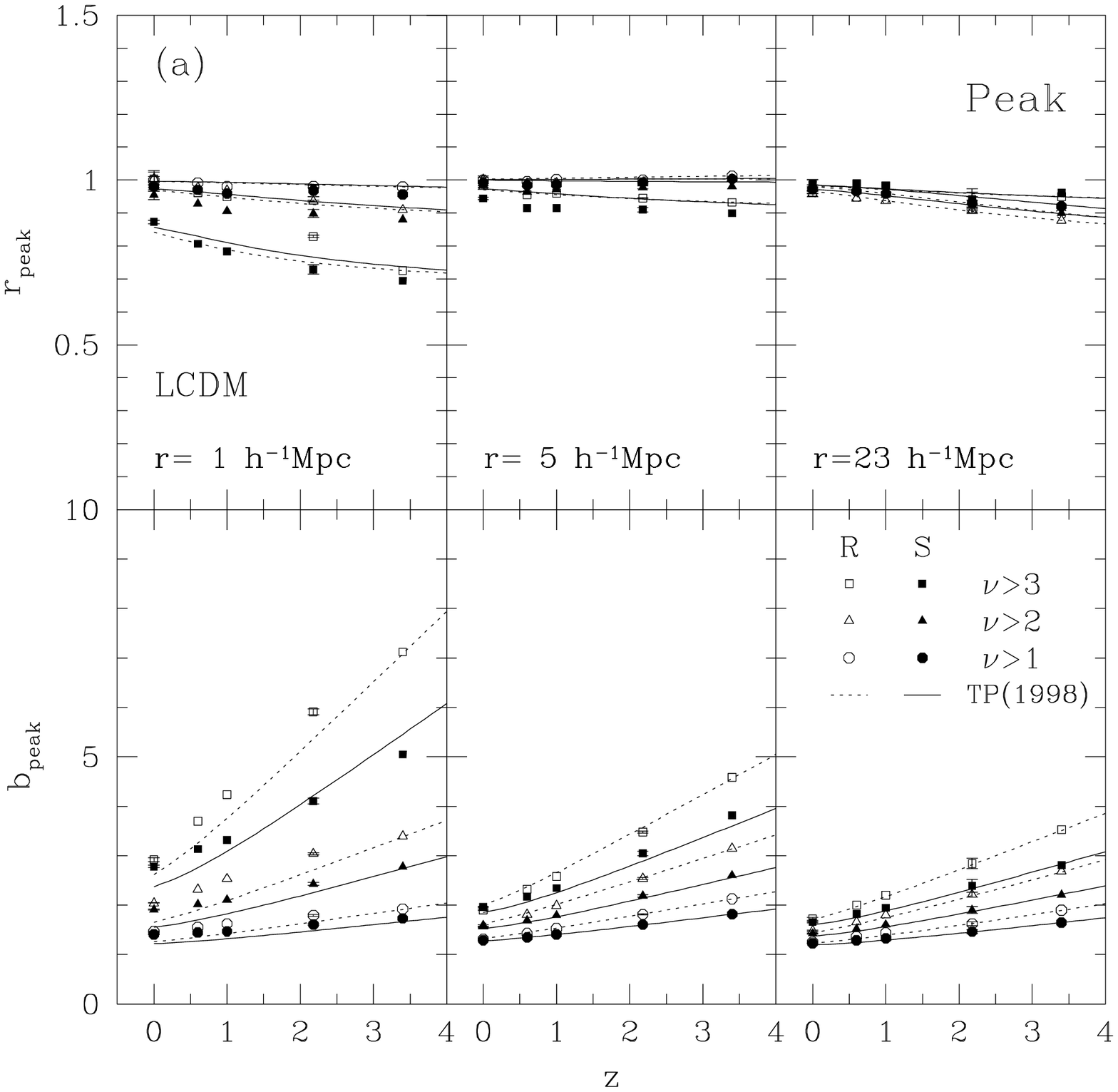}
   \leavevmode\epsfxsize=0.48\textwidth \epsfbox{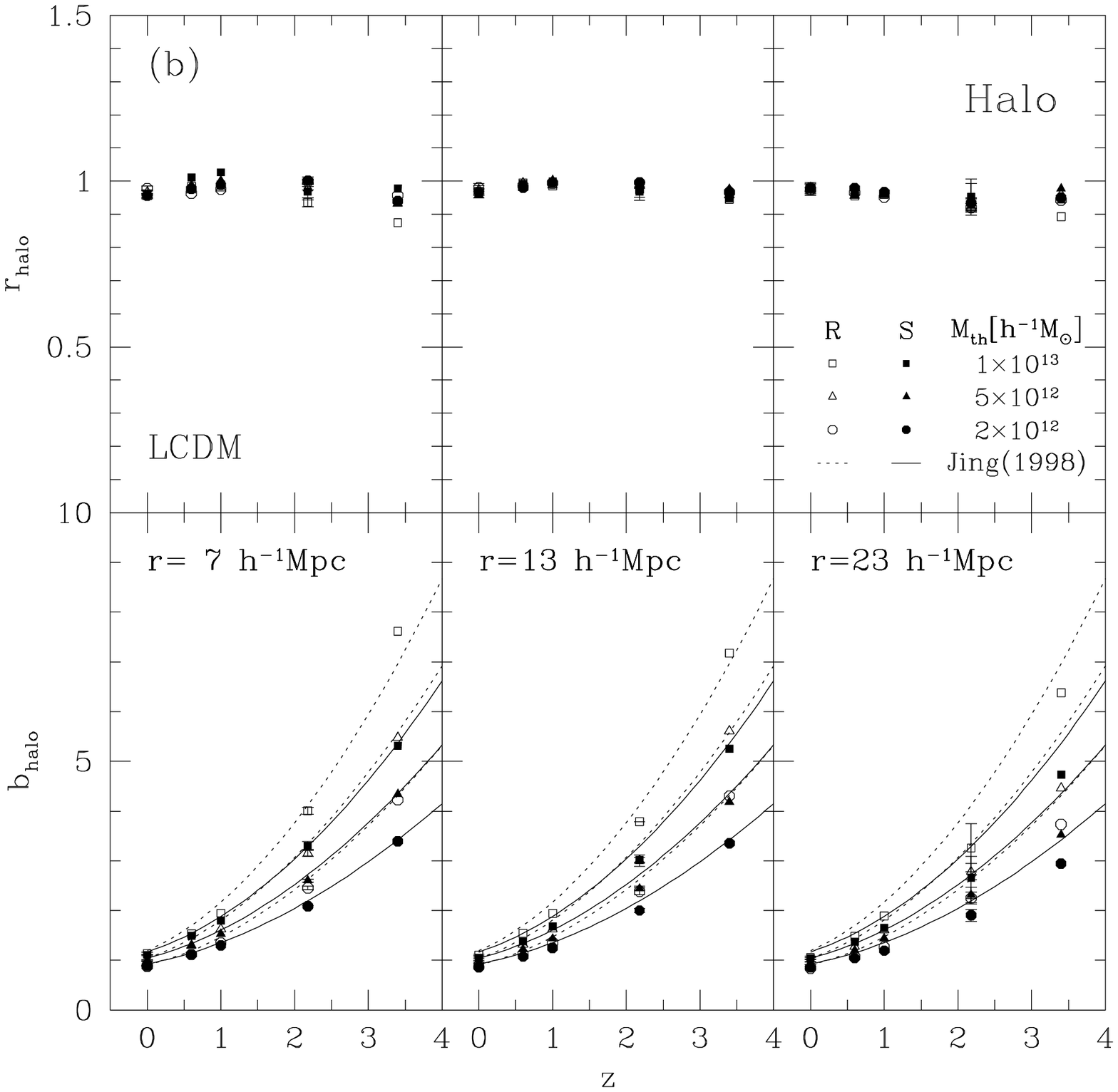}
\caption{Time evolution of the biasing and cross-correlation factors 
  in LCDM model.  (a) Open (filled) symbols indicate the peak
  simulation results for $\nu>3$ (squares), $\nu>2$ (triangles), and
  $\nu>1$ (circles) peaks in real (redshift) spaces.  Dotted and solid
  lines correspond to the model predictions of Tegmark \& Peebles
  (1998) in real and redshift spaces, respectively, normalized by the
  data at $z=3.4$.  (b) Open (filled) symbols indicate the halo
  simulation results for $M > 10^{13}h^{-1}M_\odot$ (squares), $M >
  5\times 10^{12}h^{-1}M_\odot$ (triangles), and $M > 2\times
  10^{12}h^{-1}M_\odot$ (circles) peaks in real (redshift) spaces.
  Dotted and solid curves correspond to the model prediction of Jing
  (1998) in real and redshift spaces, respectively. (b) halo model.  }
\end{center}
\end{figure}

\begin{figure}[ht]
\begin{center}
   \leavevmode\epsfxsize=0.48\textwidth \epsfbox{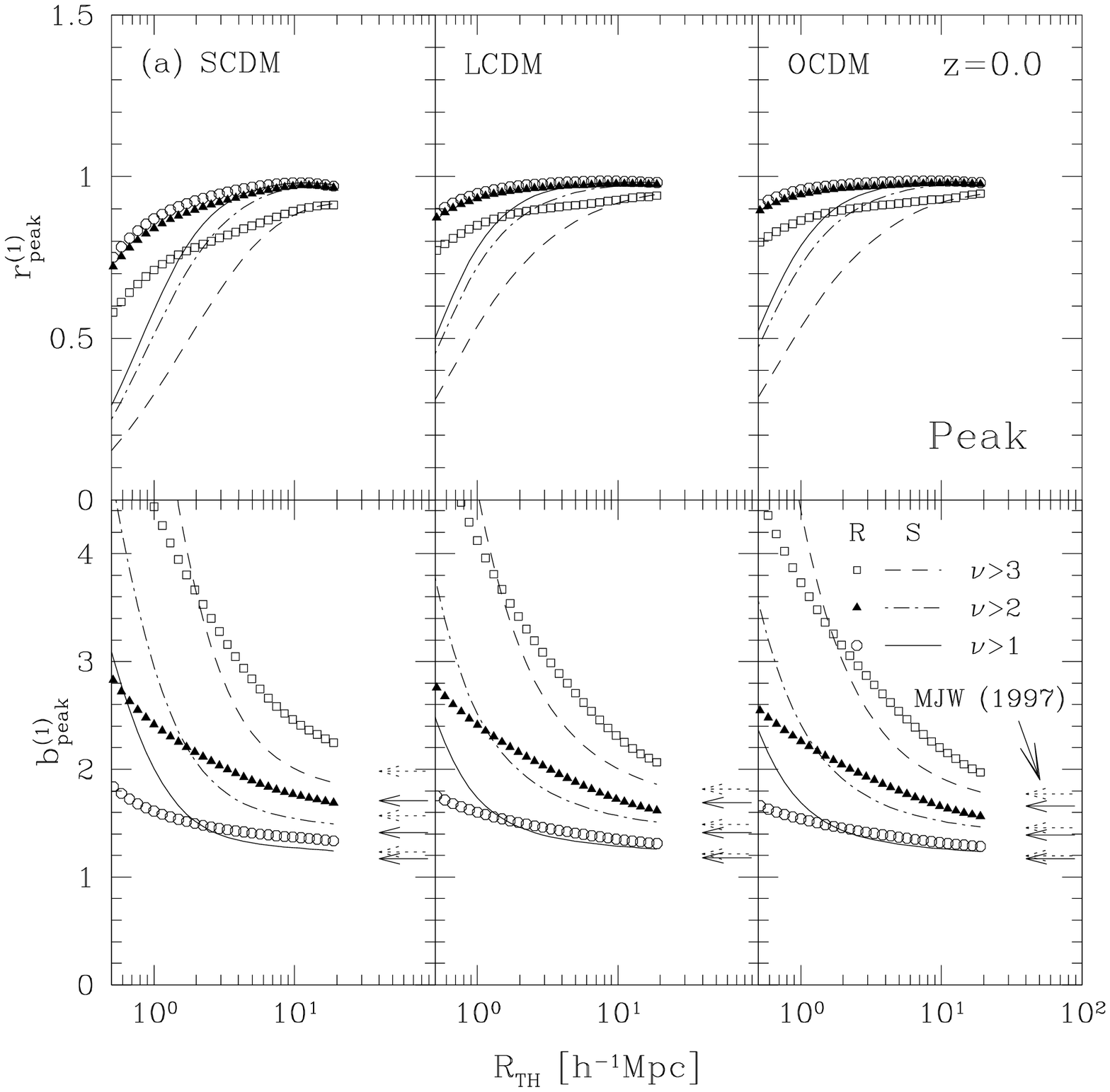}
   \leavevmode\epsfxsize=0.48\textwidth \epsfbox{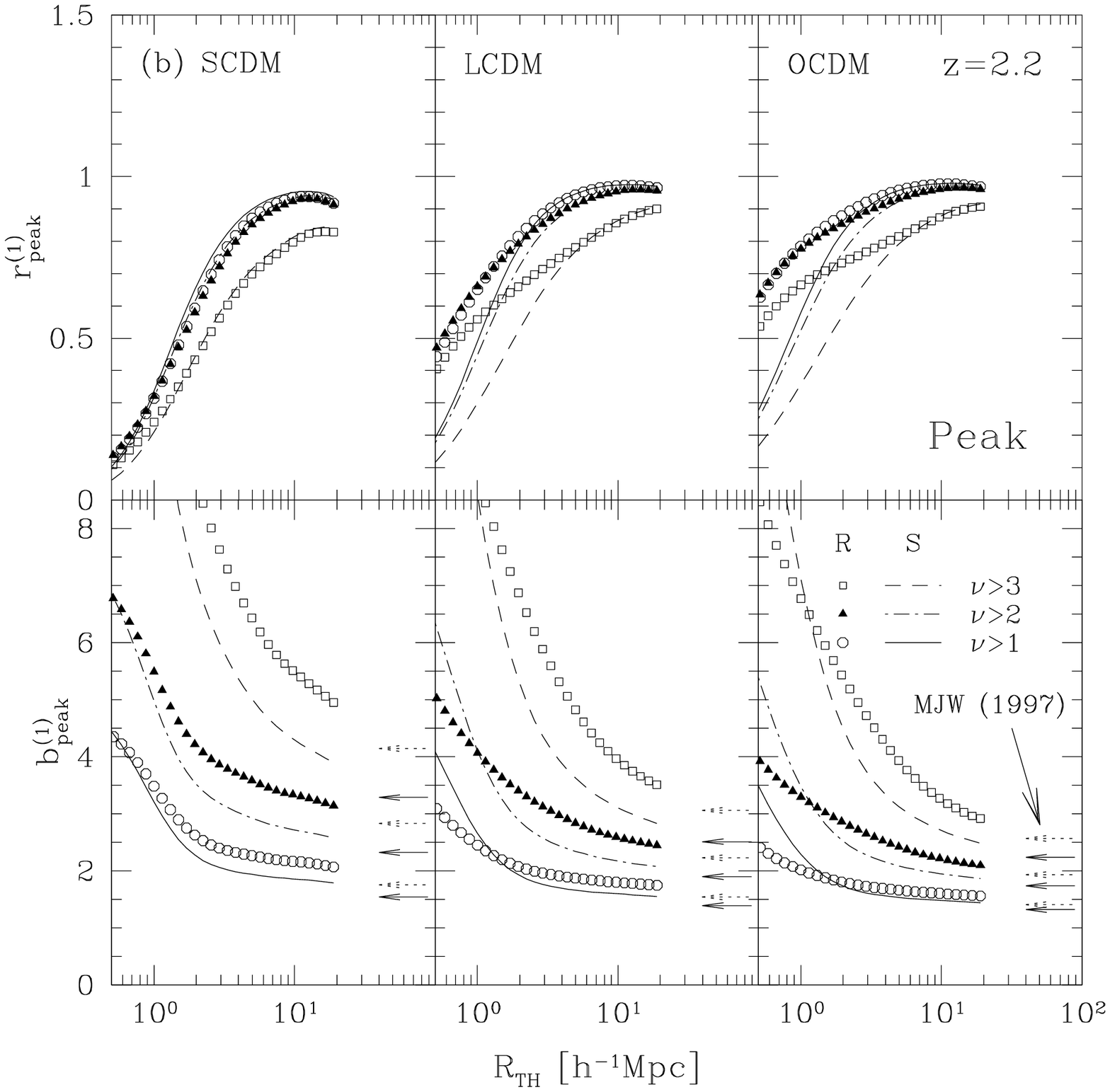}
\caption{Scale-dependence of the biasing and cross-correlation
  factors for one-point statistics of peak model. 
  Different symbols and lines have the same meaning as in Fig.2. 
  The arrows in the bottom panels indicate the 
  same model predictions as in Fig.4. 
(a) $z=0$, (b) $z=2.2$. }
\end{center}
\end{figure}
\begin{figure}[ht]
\begin{center}
   \leavevmode\epsfxsize=0.48\textwidth \epsfbox{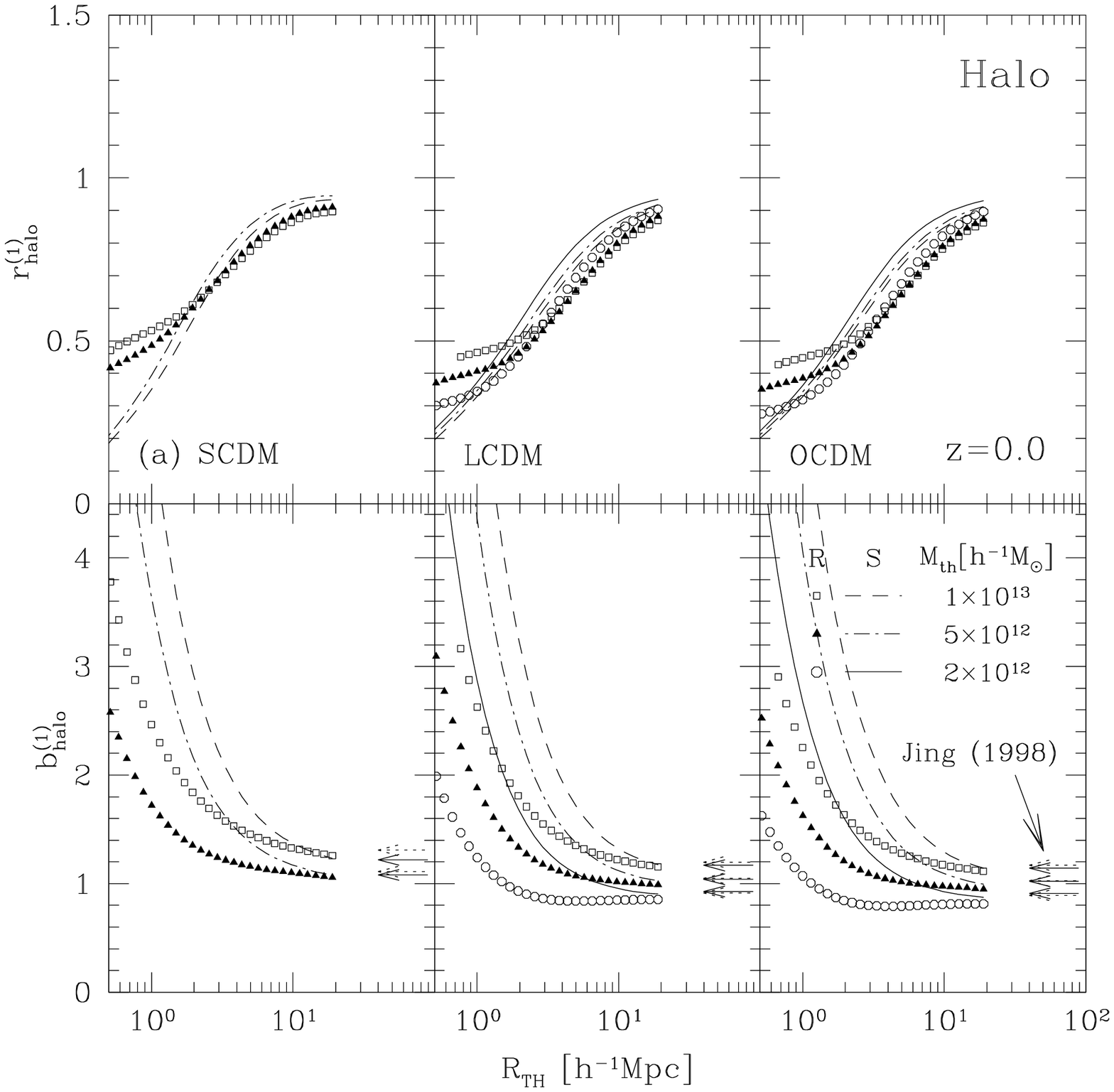}
   \leavevmode\epsfxsize=0.48\textwidth \epsfbox{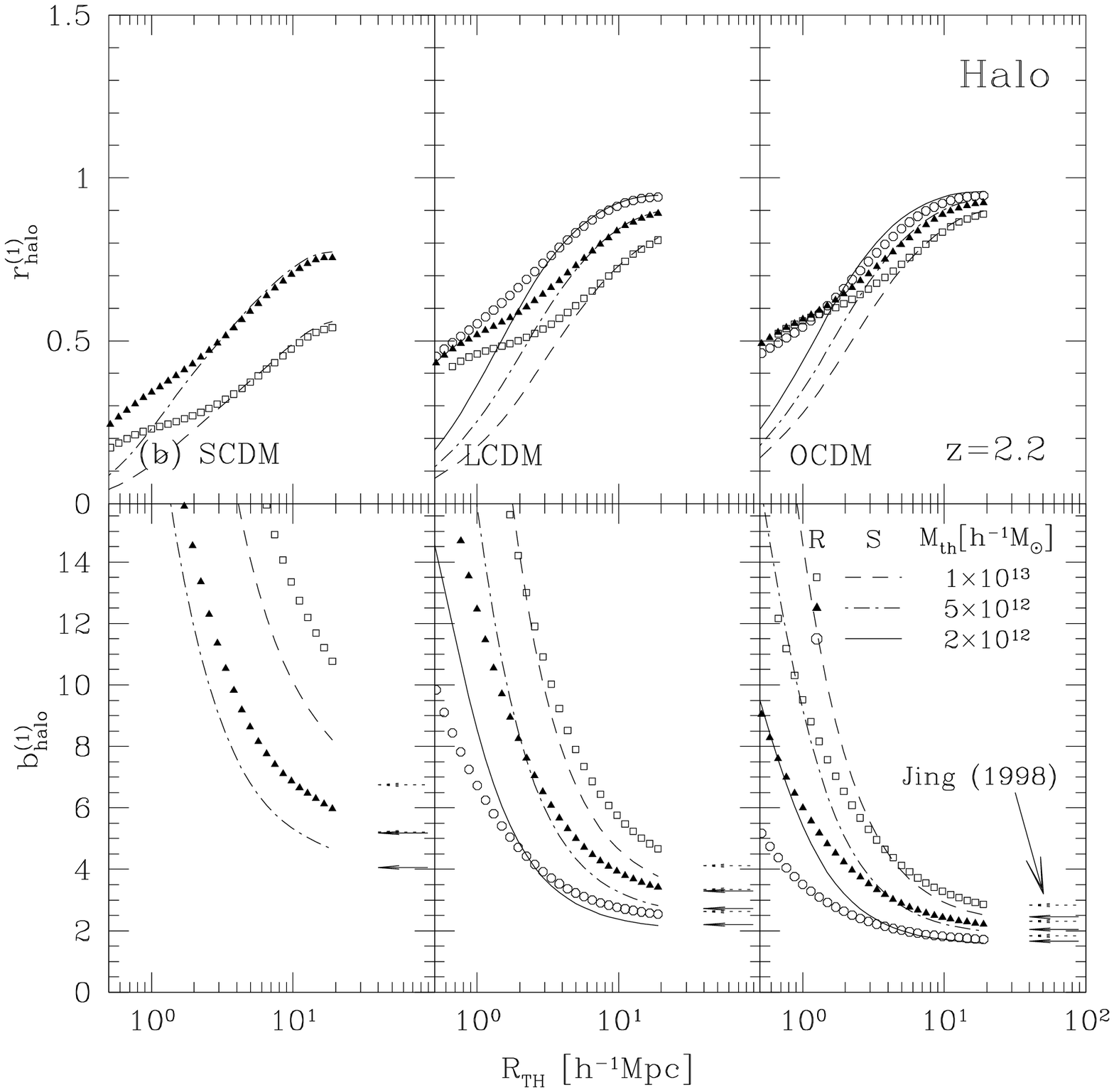}
\caption{Same as FIg. 7, but for halo model. 
  Different symbols and lines have the same meaning as in Fig.3. 
  The arrows in the bottom panels indicate the 
  same model predictions as in Fig.5. 
(a) $z=0$, (b) $z=2.2$. }
\end{center}
\end{figure}

\begin{figure}[ht]
\begin{center}
   \leavevmode\epsfxsize=0.5\textwidth \epsfbox{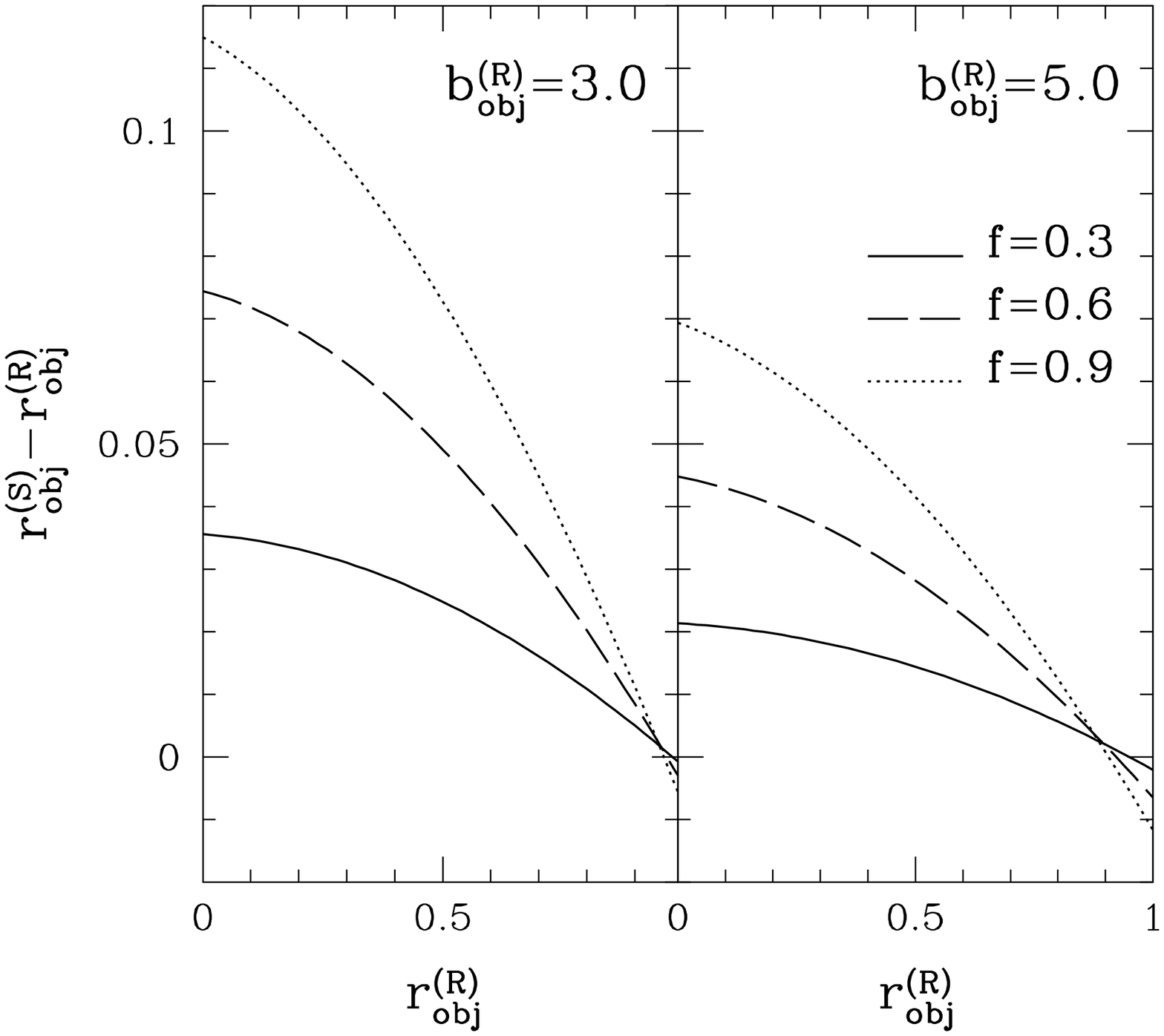}
\caption{ 
  The difference $r^{(S)}_{\rm \scriptscriptstyle obj}-r^{(R)}_{\rm
    \scriptscriptstyle obj}$ as a function of $r^{(R)}_{\rm
    \scriptscriptstyle obj}$ for $f=0.3$ (solid), $f=0.6$ (dashed),
  and $0.9$ (dotted) with $b^{(R)}_{\rm \scriptscriptstyle obj}=3.0$
  ({\it left-panel}) and $b^{(R)}_{\rm \scriptscriptstyle obj}=5.0$
  ({\it right-panel}).
 }
\end{center}
\end{figure}

\label{last}

\end{document}